\documentclass[sigconf]{acmart}

\renewcommand\footnotetextcopyrightpermission[1]{} % removes footnote with conference information in first column
\usepackage{framed,multirow}
\usepackage{url}
\usepackage{xcolor}
\usepackage{hyperref}
\usepackage{ragged2e}

\usepackage{pifont}
\newcommand{\cmark}{\ding{51}}
\newcommand{\xmark}{\ding{55}}

\AtBeginDocument{%
  \providecommand\BibTeX{{%
    \normalfont B\kern-0.5em{\scshape i\kern-0.25em b}\kern-0.8em\TeX}}}

\setcopyright{acmcopyright}
\copyrightyear{2021}
\acmYear{2021}
\acmDOI{}

\acmConference[]{}{}{}
\acmBooktitle{}
\acmPrice{}
\acmISBN{}

\acmSubmissionID{1020}

\begin{document}

%%
%% The "title" command has an optional parameter,
%% allowing the author to define a "short title" to be used in page headers.
\title{A Two-step Surface-based 3D Deep Learning Pipeline for Segmentation of Intracranial Aneurysms}

%%
%% The "author" command and its associated commands are used to define
%% the authors and their affiliations.
%% Of note is the shared affiliation of the first two authors, and the
%% "authornote" and "authornotemark" commands
%% used to denote shared contribution to the research.

\author{Xi Yang}
\affiliation{
%   \institution{Graduate School of Information Science and Technology, The University of Tokyo}}
  \institution{The University of Tokyo}}
%\ead{earthyangxi@gmail.com}
\author{Ding Xia}
\affiliation{
%   \institution{Graduate School of Information Science and Technology, The University of Tokyo}}
  \institution{The University of Tokyo}}
\author{Taichi Kin}
\affiliation{
%   \institution{The University of Tokyo Hospital, The University of Tokyo}}
  \institution{The University of Tokyo}}
\author{Takeo Igarashi}
% \ead{takeo@acm.org}
\affiliation{
%   \institution{Graduate School of Information Science and Technology, The University of Tokyo}}
  \institution{The University of Tokyo}}

% \author{Ben Trovato}
% \authornote{Both authors contributed equally to this research.}
% \email{trovato@corporation.com}
% \orcid{1234-5678-9012}
% \author{G.K.M. Tobin}
% \authornotemark[1]
% \email{webmaster@marysville-ohio.com}
% \affiliation{%
%   \institution{Institute for Clarity in Documentation}
%   \streetaddress{P.O. Box 1212}
%   \city{Dublin}
%   \state{Ohio}
%   \country{USA}
%   \postcode{43017-6221}
% }

% \author{Lars Th{\o}rv{\"a}ld}
% \affiliation{%
%   \institution{The Th{\o}rv{\"a}ld Group}
%   \streetaddress{1 Th{\o}rv{\"a}ld Circle}
%   \city{Hekla}
%   \country{Iceland}}
% \email{larst@affiliation.org}

%%
%% By default, the full list of authors will be used in the page
%% headers. Often, this list is too long, and will overlap
%% other information printed in the page headers. This command allows
%% the author to define a more concise list
%% of authors' names for this purpose.
\renewcommand{\shortauthors}{Yang et al.}

%%
%% The abstract is a short summary of the work to be presented in the
%% article.
\begin{abstract}
The exact shape of intracranial aneurysms is critical in medical diagnosis and surgical planning. While voxel-based deep learning frameworks have been proposed for this segmentation task, their performance remains limited. In this study, we offer a two-step surface-based deep learning pipeline that achieves significantly higher performance. Our proposed model takes a surface model of entire principal brain arteries containing aneurysms as input and returns aneurysms surfaces as output. A user first generates a surface model by manually specifying multiple thresholds for time-of-flight magnetic resonance angiography images. The system then samples small surface fragments from the entire brain arteries and classifies the surface fragments according to whether aneurysms are present using a point-based deep learning network (PointNet++). Finally, the system applies surface segmentation (SO-Net) to surface fragments containing aneurysms. We conduct a direct comparison of segmentation performance by counting voxels between the proposed surface-based framework and the existing voxel-based method, in which our framework achieves a much higher dice similarity coefficient score ($72\%$) than the prior approach ($46\%$).
\end{abstract}

%%
%% The code below is generated by the tool at http://dl.acm.org/ccs.cfm.
%% Please copy and paste the code instead of the example below.
%%
\begin{CCSXML}
<ccs2012>
   <concept>
       <concept_id>10010405.10010444.10010449</concept_id>
       <concept_desc>Applied computing~Health informatics</concept_desc>
       <concept_significance>500</concept_significance>
       </concept>
   <concept>
       <concept_id>10010147.10010178.10010224.10010245</concept_id>
       <concept_desc>Computing methodologies~Computer vision problems</concept_desc>
       <concept_significance>500</concept_significance>
       </concept>
 </ccs2012>
\end{CCSXML}

\ccsdesc[500]{Applied computing~Health informatics}
\ccsdesc[500]{Computing methodologies~Computer vision problems}

%%
%% Keywords. The author(s) should pick words that accurately describe
%% the work being presented. Separate the keywords with commas.
\keywords{Intracranial aneurysm segmentation, Point-based 3D deep learning, Medical image segmentation}

%% A "teaser" image appears between the author and affiliation
%% information and the body of the document, and typically spans the
%% page.
% \begin{teaserfigure}
%   \includegraphics[width=\textwidth]{sampleteaser}
%   \caption{Seattle Mariners at Spring Training, 2010.}
%   \Description{Enjoying the baseball game from the third-base
%   seats. Ichiro Suzuki preparing to bat.}
%   \label{fig:teaser}
% \end{teaserfigure}

%%
%% This command processes the author and affiliation and title
%% information and builds the first part of the formatted document.
\maketitle

\section{Introduction}
An intracranial aneurysm (IA) is a weakened or thinned portion of a blood vessel in the brain that bulges dangerously and fills up with blood. Bloated aneurysms compress the surrounding nerves and brain tissue and have a high risk of rupture, resulting in subarachnoid hemorrhage (SAH). The risk of such rupture is related to the size and form of the IA~\cite{ucas2012natural}. The practical surgical approach is to clip their neck to prevent the rupture. Therefore, extracting the shape of aneurysms is a crucial aspect not only of IA diagnosis but also of preoperative examination to determine the position and posture of the necessary clips~\cite{alaraj2015virtual}. In current practice, 
%this process is manual, and draws the experience of experts; each case requires several minutes.
this process requires manual identification by medical experts, taking several minutes per case. Clearly, automating this process is a very worthwhile venture. Furthermore, employing automation, we can also obtain large segmented datasets, which can open up new avenues for research toward gaining further insights into IA through statistical analysis. 

Over the last decade, many extraction algorithms have been designed by calculating local geometric features~\cite{nikravanshalmani2010three, law2012segmentation}; however, rule-based methods have high computational costs and time requirements, and their performance is limited because of the wide variety of aneurysm shapes. Meanwhile, deep learning techniques are becoming increasingly popular in medical image processing; 
%however, few studies were published on segmentation of IAs (\cite{sichtermann2019deep, yang2020intra}). In addition, we describe the details of state-of-the-art related works in Sec.~\ref{sec:related_works}. 
however, they are mostly used for classification and detection. Few prior research works have explored the application of deep learning methods to the segmentation of IAs, and their performance remains limited~\cite{nikravanshalmani2010three} (see Section~\ref{sec:related_works}).

\begin{table}[t]
\caption{Comparison of related works and our method.}
\label{tab:comparison_m}
\centering
\begin{tabular}{c|c|c}
\hline & Entire image & Surface-based \\
       & (Practicality) & (Effectiveness) \\
\hline Park et al.~\cite{park2019deep} \& & \multirow{2}{*}{\cmark} & \multirow{2}{*}{\xmark} \\
       Sichterman et al.~\cite{sichtermann2019deep} & & \\
\hline Yang et al.~\cite{yang2020intra} \& & \multirow{2}{*}{\xmark} & \multirow{2}{*}{\cmark} \\
       Bizjak et al.~\cite{bizjak2020vascular} & & \\
\hline Our & \cmark & \cmark \\
\hline
\end{tabular}
\end{table}

\begin{figure}[t]
\centering
\includegraphics[width=0.38\textwidth]{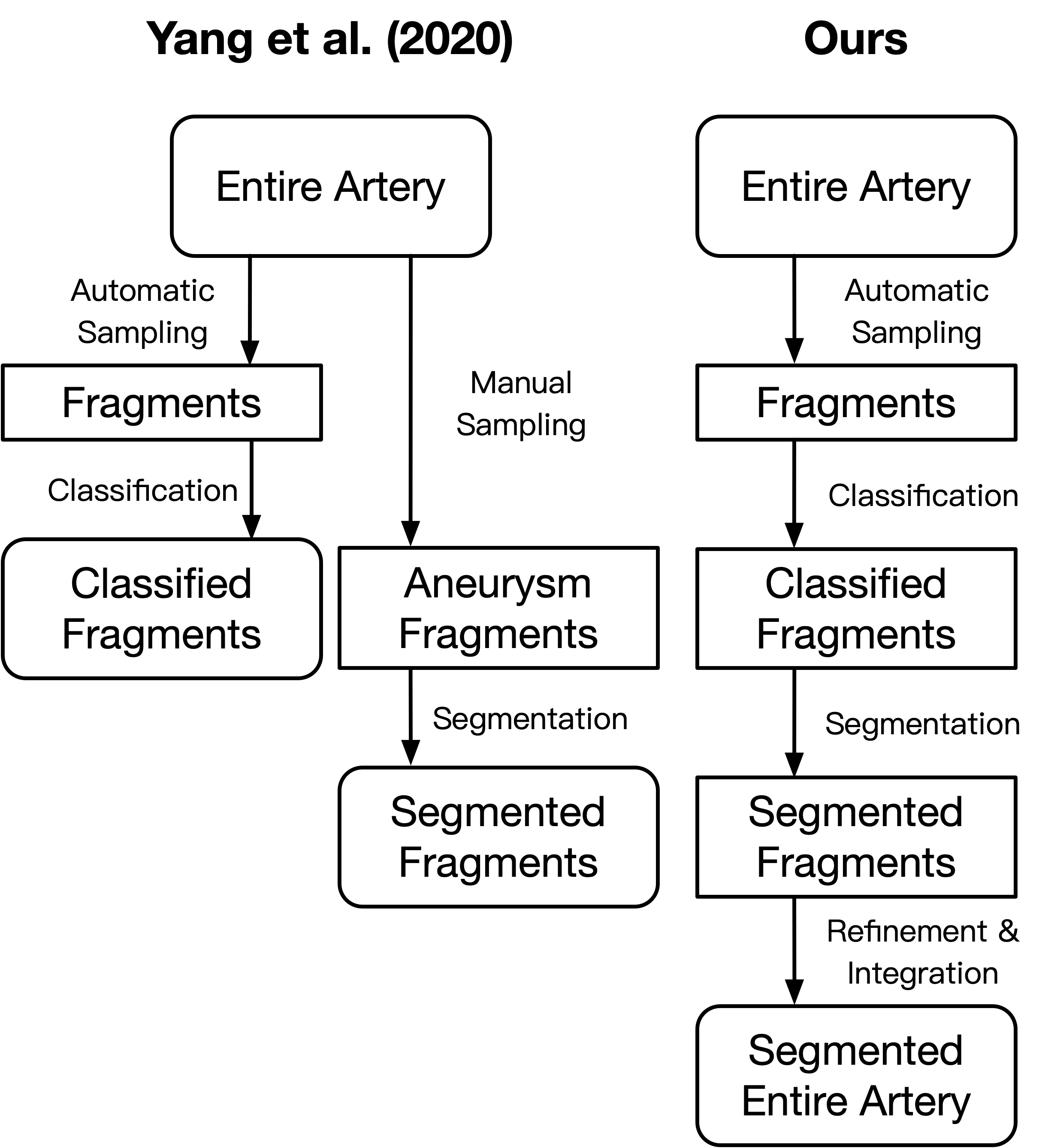}
\caption{Comparison of the pipeline of Yang et al.~\cite{yang2020intra} and Ours.}
\label{fig:pipeline}
\end{figure}

This study builds on Yang et al.'s work %!!!our previous work, 
the IntrA dataset~\cite{yang2020intra}, which was created for surface-based classification and segmentation of IAs, and reported the performance of existing neural network models on both tasks. However, in their work, %!!!in the previous work, 
the dataset and execution process were fully separated for both classification and segmentation tasks. Segmentation was evaluated only on manually sampled surface fragments containing aneurysms. This process is unrealistic in clinical practice. In addition, the per-fragment segmentation results were not integrated. Therefore, in this study, we present a complete processing pipeline for segmenting IAs, as shown in Figure~\ref{fig:pipeline}, by integrating deep learning and geometry processing techniques to achieve better performance. Our proposed pipeline takes an entire intracranial vessel network model as input and returns IA fragments as output.

The main contributions of this study are as follows:
\begin{enumerate}
\item We propose a complete pipeline using point-based 3D deep neural networks for aneurysm segmentation from entire medical images. The proposed pipeline with automatic sampling achieves SOTA results comparable to segmentation based on manual sampling~\cite{yang2020intra}. %\textcolor{red}{(Because one reviewer of MICCAI questioned that the final performance of this study is not improved comparing with the results of our CVPR paper. I want to clarify that our inputs are different. This study is harder than CVPR paper, but we obtained same level final performance.)}
% \item We compared the experimental results between our two-steps design and segmentation only method on our dataset by 5-fold cross validation. 
\item We demonstrate the advantage of our two-step pipeline combining a classification step and a segmentation step by comparing it against a segmentation-only pipeline.
% \item We also present solid comparison experiment results, especially the results of a head-to-head comparison between our surface-based method and a SOTA volume-based method in a segmentation task performed on entire brains~(\cite{podgorsak2019automatic}).
\item We present a direct comparison between our surface-based framework and a SOTA voxel-based method, showing the superiority of our proposed framework.
\end{enumerate}

% \textbf{Our proposed pipeline} In this study, we first semi-automatically obtained the surface representation of the entire intracranial blood vessel. We then collected small fragments from the entire vessel network and performed surface-based classification on them. Finally, we applied surface-based segmentation to the fragments that were classified as those containing aneurysms. 

\section{Related works}
\label{sec:related_works}

In the field of computer vision in general, detection tasks usually involve determining the rough location and size of target objects, for example, a boundary box, whereas, in medical imaging, detection may indicate only the presence of target objects. The difficulties of both are lower than those of the segmentation task, which requires algorithms to predict the precise shape of the target objects.

\subsection{Detection}
Deep learning methods have been widely used as diagnosis aid to detect IAs. %We introduce several SOTA methods to distinguish the difference with segmentation task.
Nakao et al.~\cite{nakao2018deep} detected intracranial aneurysm-based MRA images using a basic deep convolutional neural network.
Ueda et al.~\cite{ueda2019deep} used ResNet-18 for an automated diagnosis of cerebral aneurysms from TOF MR angiography image data from several sources.
% \cite{jerman2015blob} proposed blob enhancement filter based on a modified volume ratio of Hessian eigenvalues to increase detection sensitivity.
In particular, Zhou et al.~\cite{zhou2019intracranial} proposed a transferable multi-model ensemble (MMEN) architecture to predict the possibility of aneurysms using a mesh model. This approach used 3D objects as input, but also still used 2D neural networks by conformal mapping.

\subsection{Segmentation}
Segmentation of IAs requires obtaining detailed location and shape information of an aneurysm. Conventional approaches have used rule-based 2D or 3D shape analyses. 
For example, Nikravanshalmani et al.~\cite{nikravanshalmani2010three, nikravanshalmani2013segmentation} used a level set algorithm and a region growing based approach for the  semi-automatic segmentation of cerebral aneurysms from CTA images. 
Law et al.~\cite{law2007vessel, law2012segmentation} proposed an intensity-based algorithm to segment intracranial vessels and embedded aneurysms using multirange filters and local variances.
Wang et al.~\cite{wang2016multilevel} presented a multilevel segmentation method based on the lattice Boltzmann method (LBM) and level set with ellipse for accurate segmentation of intracranial aneurysms. 
Sulayman et al.~\cite{sulayman2016semi} proposed a scheme for semi-automatic detection and segmentation of intracranial aneurysms.
Dakua et al.~\cite{dakua2018pca} presented a PCA-based approach to segmenting the brain vasculature in low contrast cerebral blood vessels.
Jerman et al.~\cite{jerman2019automated} proposed automated cutting plane (ACP) positioning, which is based on the detection of specific geometric descriptors of an aneurysm and its parent vasculature.

Recently, learning-based methods have become increasingly popular alongside the development of deep learning. However, few studies have focused on the segmentation of IAs. 
Podgorsak et al.~\cite{podgorsak2019automatic} claimed that segmenting IAs and the surrounding vasculature from digital subtraction angiography (DSA) images using a convolutional neural network was non-inferior to manually identifying the contours of aneurysms. However, they extracted only 2D contours of the IAs.
Park et al.~\cite{park2019deep} developed a neural network segmentation model called HeadXNet to generate voxel-by-voxel predictions of intracranial aneurysms on tomographic angiography (CTA) imaging to augment the performance of clinical intracranial aneurysm diagnosis. However, they evaluated their model based on the sensitivity, specificity, and accuracy of the entire image; these metrics cannot reflect actual segmentation performance in practice.
Sichtermann et al.~\cite{sichtermann2019deep} applied a popular software-based on a volume-based neural network, called DeepMedic~\cite{kamnitsas2017efficient}, to segment IAs from MRA images. However, the performance of their approach was suboptimal ($46\%$ in DSC).
Importantly, Yang et al.~\cite{yang2020intra} and Bizjak et al.~\cite{bizjak2020vascular} made useful attempts to apply point-based networks to segment IAs. However, segmentation was only performed for surface fragments that were manually labeled as containing aneurysms. This process is unrealistic in clinical practice. %In Addition, in the work \cite{bizjak2020vascular}, their dataset division may not solid, the evaluation results on testset and val set are higher than train set. It means the hard cases are only divided into train set. And they also only employed the sensitivity of entire image as evaluation metric.
In addition, Bizjak et al.~\cite{bizjak2020vascular} employed only the sensitivity of the entire input as an evaluation metric.

\subsection{3D deep learning}

3D surface models have several representations, including projected view, voxel/pixel, point cloud, and mesh. Voxel-based deep learning approaches are easy to implement using networks developed for 2D image tasks. However, point-based methods have shown great promise and improved performance compared with previous voxel-based methods in the field of 3D shape analysis using deep learning~\cite{qi2017pointnet}. In addition, using point-based rather than mesh-based methods~\cite{hanocka2019meshcnn} avoids arduous pre-processing steps, including cleaning the models and constructing manifold meshes. %Unfortunately, the point-based methods can not used on the analysis of medical images directly. The publicly available dataset IntrA~(\cite{yang2020intra}) introduced a manually created 3D surface model dataset to provide a solution of this problem.
A problem with point-based methods is that they require surface models and cannot be directly applied to medical images. Therefore, we introduce an interactive surface reconstruction process before applying point-based classification and segmentation. We leverage Yang et al.~\cite{yang2020intra}'s surface model data set for training and evaluation.

\section{Proposed Pipeline}

\begin{figure*}[t]
\includegraphics[width=\textwidth]{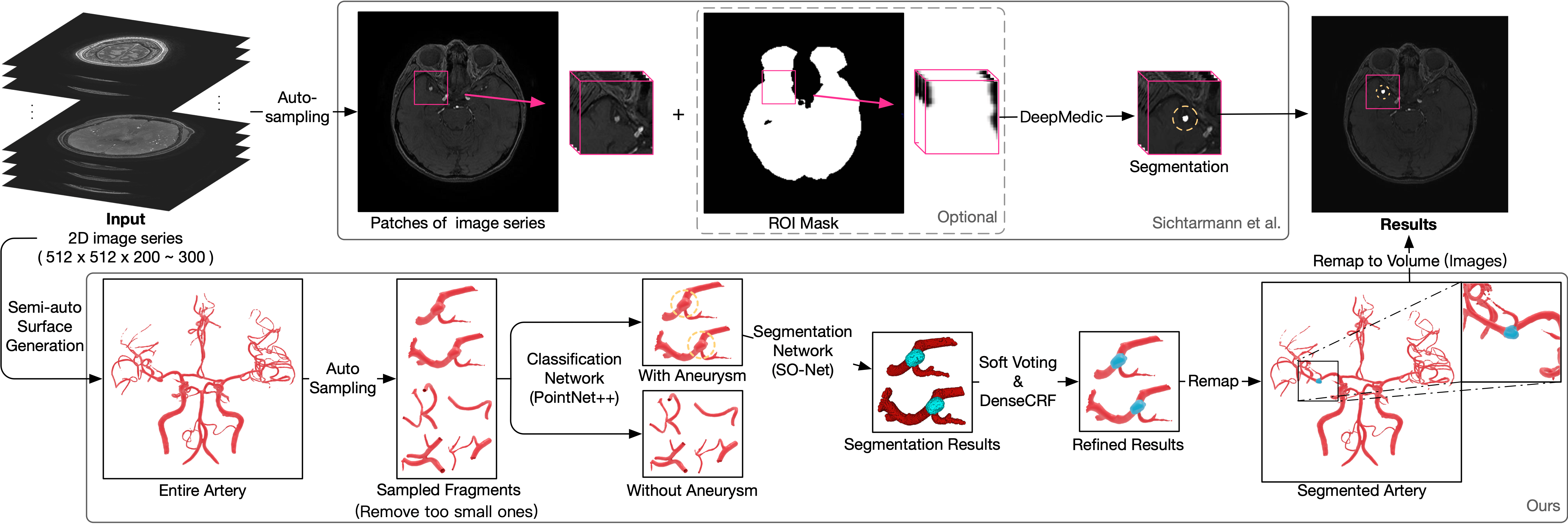}
% \caption{Comparison of the proposed framework and the existing approaches. The framework of pixel/voxel-based methods (Sichtermann et al.) is simple, but has an unsatisfactory segmentation accuracy. Yang et al.'s dataset verified the performance of point/mesh-based networks on medical images, but could not directly applied on clinical application. (remove middle??}
\caption{Comparison of our proposed pipeline and voxel-based method.} %The approach of pixel/voxel-based methods~\cite{sichtermann2019deep} is simple, but has an unsatisfactory segmentation accuracy.}
\label{fig:framework}
\end{figure*}

Figure~\ref{fig:framework} compares our surface-based pipeline with a voxel-based method. In the voxel-based method~\cite{sichtermann2019deep}, the medical image is directly fed to a neural network, which affixes a label to each voxel indicating whether it is part of an aneurysm or not. 

Our surface-based pipeline first interactively reconstructed a surface model of the entire principal brain arteries using a multiple threshold method. We then generated small samples along vessels within the entire model and performed surface-based classification (PointNet++) on them. Finally, we performed surface-based segmentation (SO-Net) on samples classified as containing aneurysms. To compare our results with the results obtained by voxel-based methods, we voxelized the surface model of the segmented aneurysms into volumes using winding-numbers~\cite{barill2018fast}.
%The new framework is improved for adapting clinical application from a series of TOF-MRA images to the reconstructed 3D shape of extracted IAs as shown in~\ref{fig:framework}. First, the 3D surface model of entire brain vessel are generated manually. Then, the segments of vessel are sampled and classified, the segments with aneurysms are inputted to the segmentation network to extract the accuracy shape of IAs. Final, the segmentation results are refined and combined as entire model. To compare with the existing method, we use winding number~\cite{barill2018fast} to generated solid models from the surface model.

% \subsection{Obtaining a Surface-Model}
\subsection{Interactive reconstruction of surface models}
We first obtained surface models of the principal brain arteries of patients using TOF-MRA image sets. We performed this semi-automatically using a software package (Amira 2019 by Thermo Fisher Scientific, MA, USA) based on a multi-threshold method~\cite{kin2012new}. Importantly, we focused on dealing with the brain regions surrounding aneurysms to ensure that the complete shape of the aneurysms was exhibited in the extracted 3D surface model, compared with the data in Intra~\cite{yang2020intra}. In the future, we envision that this process can be mostly or fully automated using a reconstruction network specifically designed for brain arteries.

%\subsection{Pre-processing}
%We briefly introduce the data processing pipeline of IntrA~\cite{yang2020intra} at first. The complete surface models are extracted from TOF-MRA, then vessels segments are automatically generated by sampling the complete models. Classification methods are verified by binary classifying these segments into with or without aneurysms. In the experiment of segmentation, they manually generate the segments of interest and restore the shape of each aneurysm. The groundtruth of aneurysms parts are annotated on these segments. Our pre-processing pipeline is shown as Figure~\ref{fig:pipeline}.

% \noindent \textbf{Surface model reconstruction.}
%\textbf{Surface model reconstruction.}
%As the paper~\cite{yang2020intra}, from a series of TOF-MRA images, a surface model of principal brain arteries of one patient is interactively reconstructed using Amira 2019 (Thermo Fisher Scientific, MA, USA) by the multi-threshold method~\cite{kin2012new}. Importantly, we focus on dealing with the surrounding region of aneurysms to ensure that the complete shape of aneurysms are exhibited in the extracted 3D surface model.

%\textbf{Annotation.} 
%Instead of annotating on aneurysm segments, we annotated the aneurysm parts on the complete model of brain arteries. Thus, both segmentation and classification labels of sampling segments could be generated automatically for deep learning networks.

% \textbf{Sampling.}
\subsection{Fragments sampling}

\begin{figure}[t]
\includegraphics[width=\linewidth]{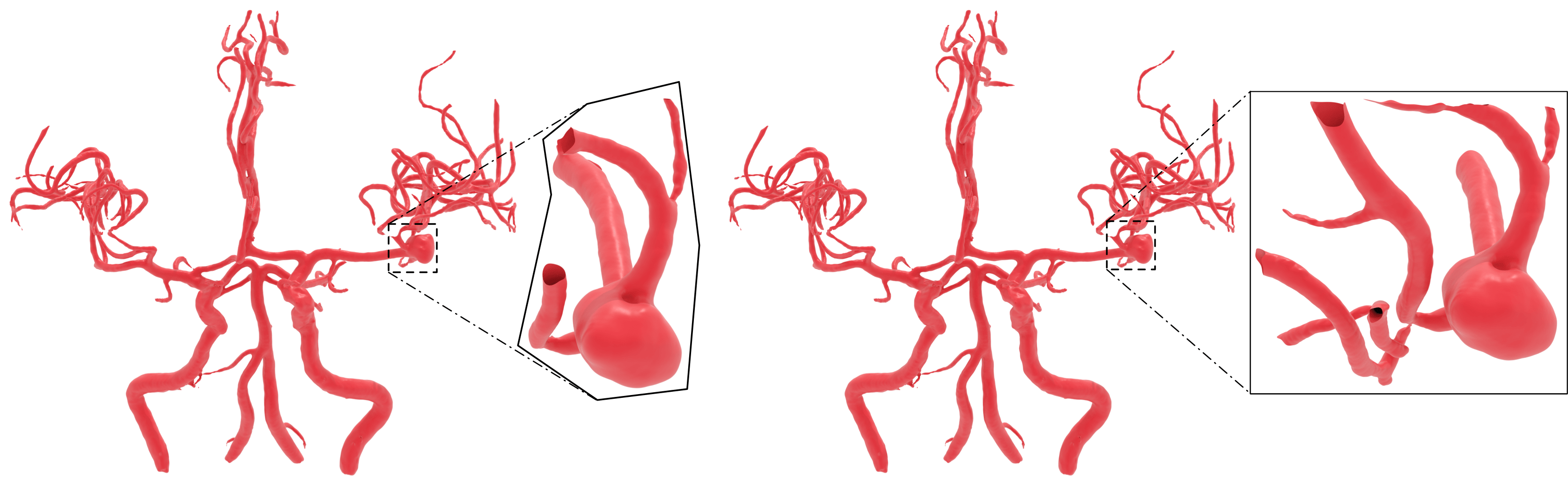}
\caption{In contrast to grid sampling, our sampling based on geodesic distance avoids noisy blood vessels are involved.}
\label{fig:sampling}
\end{figure}

The segmentation network does not work well if the entire model is input directly because the aneurysm portions are tiny compared to the entire model. Therefore, we first sampled small fragments from the entire model. %(The size of fragments should be slightly larger than aneurysms.!!!) 
%Grid box is a normal method to obtain small patches in medical image processing, however, it is hard to obtain clean artery surfaces. 
A typical method used to obtain small patches in medical images involves systematically sampling rectilinear boxes; however, it is difficult to obtain clean artery surfaces because surface boundaries are not aligned with canonical axes. Thus, we designed a sampling algorithm along a surface, as shown in Figure~\ref{fig:sampling}.
We first set the size of the fragments such that they roughly covered an aneurysm of typical size according to the experience of the medical experts. 
%It is roughly covers an aneurysm of usual size. 
Then, we divided the 3D space into regular grid cells. From the center of each grid cell, the nearest point on the surface model closer than a threshold ($\alpha$) was selected as a starting point, while grid cells that did not have nearby surface model points were ignored. Finally, we collected the surface points around the starting points whose geodesic distance was less than a threshold ($\beta$). Note that this sampling was designed to cover the model with some overlap; thus, uniform sampling is not a vital requirement of our proposed method.

\begin{figure*}[t]
\includegraphics[width=\textwidth]{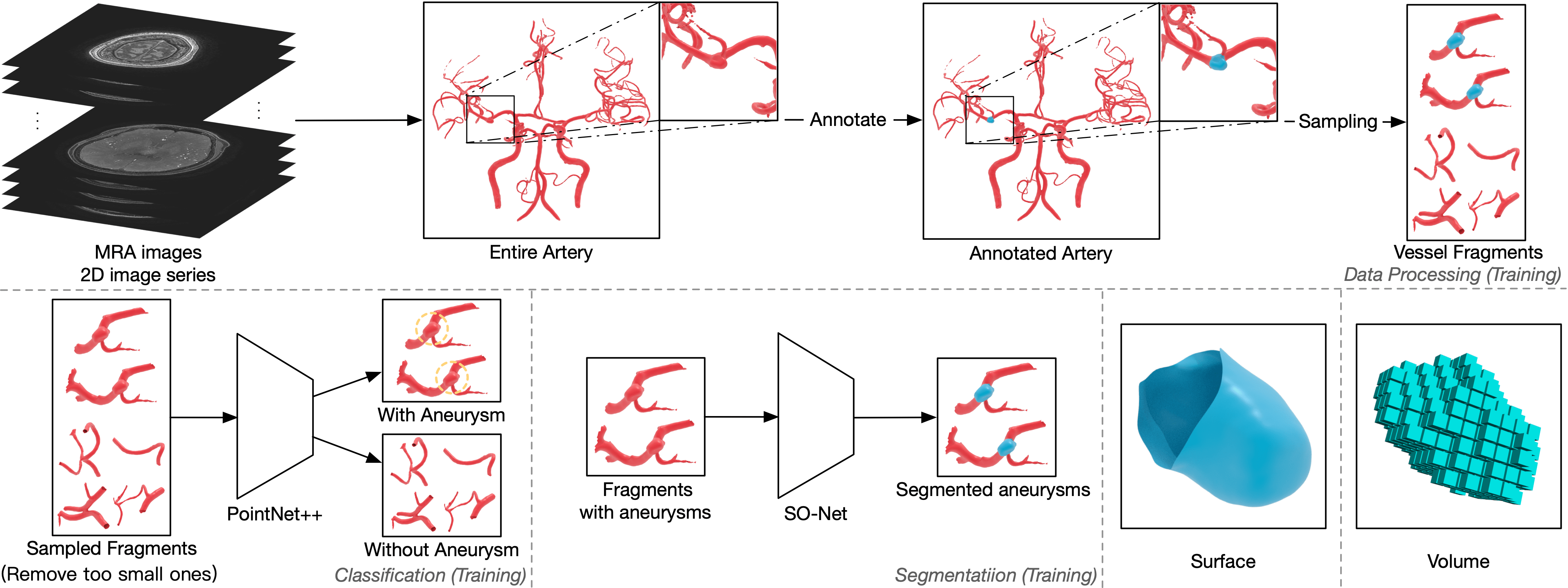}
\caption{Details of our data-processing pipeline and algorithm.} 
\label{fig:details}
\end{figure*}

% \subsection{Algorithm}
% Considering the almost segments do not include aneurysms, we use two-stages, classification-segmentation, in our framework for efficiency. We first use PointNet++~\cite{qi2017pointnet++} to classify the segments into two classes, with or without aneurysms. The segments with too small number of points are ignored before classification. Then, the segmentation network, SO-Net~\cite{li2018so}, only inputs the segments with aneurysms to extract aneurysm parts. During test, the input of segmentation network is the segments which are predicted into the aneurysm class by classification network. These two point-based network are selected according to the benchmark of IntrA. For better applicability, we avoid using mesh-based networks to omit the mesh clean step. 

\subsection{Classification step}
We used PointNet++~\cite{qi2017pointnet++} to classify the fragments into two classes, distinguishing those with and without aneurysms. Fragments with few points were discarded before classification. However, the number of fragments with aneurysms was still significantly fewer than those without aneurysms. Therefore, we used a weighted soft-max cross-entropy loss function to train the classification network to deal with the imbalance between the two classes. The purpose of the classification step was to reduce the number of candidate fragments fed to the segmentation network and improve its accuracy.

By design, we sacrifice some classification performance to obtain a better segmentation result. See Section~\ref{sec:results} for a detailed discussion. The evaluation of our classification results was not fully equal to the accuracy of the detection task. Our sampling method allows one IA to be sampled into several fragments, and fragments with a tiny portion of IA may be misclassified, but the same IA could be detected from other fragments. Therefore, the real-world detection performance results are much better than the performance of the classifier itself.

We expected the classification step to reduce training and prediction time as well as data noise in the final segmentation result. To demonstrate this, we conducted an experiment comparing results between the proposed two-step pipeline and a pipeline without the classification step (Section~\ref{subsec:seg_only}).

\subsection{Segmentation step}
We then fed fragments with aneurysms into the segmentation network, SO-Net~\cite{li2018so}. Only a fraction of the original points were classified after segmentation because the point-based network uses random sampling to deal with input models with varying numbers of points. Thus, we performed segmentation with random sampling multiple times and assigned labels to all the points based on a voting criterion to enrich segmentation details. There was the possibility of a small number of points failing to be sampled; we labeled them arteries rather than aneurysms; however, these were few, and did not significantly affect the segmentation results. Next, we used a conditional random field (CRF) to refine each voting result, specifically DenseCRF~\cite{krahenbuhl2011efficient}. Finally, the segmentation results of the individual fragments were remapped into the original entire model using a global ID for each point to obtain a complete segmentation result over the entire surface. Points in overlapped parts are commonly sampled twice, so we did not use majority voting. Points with multiple labels were marked as aneurysms if they had an aneurysm label. 

\subsection{Voxelization}
We converted the results of our surface-based segmentation to volume to perform a direct comparison with voxel-based methods. An example is shown in Figure~\ref{fig:details}. We first obtained a set of the query points through uniform sampling using the same interval as the MRA images. We then computed the winding number of each query point using the fast winding number method~\cite{barill2018fast} to determine whether a given point was inside or outside an aneurysm. We set the threshold of the winding number to $0.5$, as suggested in their study. This step is not necessary for clinical practice if segmentation results are required only on the surface model.

\section{Experiments}
Various medical imaging techniques, such as computed tomography angiography (CTA), magnetic resonance angiography (MRA), and digital subtraction angiography (DSA) can be used to obtain images of the brain. DSA is the most sensitive method for diagnosing intracranial IAs; however, it is invasive and time-consuming. Although CTA scans are efficient, distinguishing the details of vessels and aneurysms using CTA remains difficult. TOF-MRA is a less invasive examination and has a high sensitivity for diagnosing IAs. Therefore, we decided upon TOF-MRA as a suitable technique for preoperative examination. However, our proposed pipeline is not affected by the type of medical image, as it is based on reconstructed surface models.

\subsection{Dataset}
\begin{figure}[t]
\includegraphics[width=\linewidth]{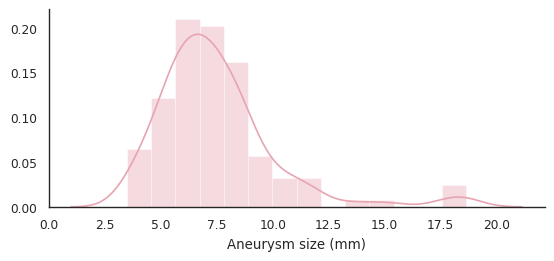}
\caption{Distribution of aneurysm sizes on our dataset. Mean: 7.49 mm, SD: 2.72 mm; Range: 3.48–18.66 mm.}
\label{fig:size}
\end{figure}

We collected TOF-MRA image sets of 103 patients with 114 aneurysms. Each set contains at least one IA, and $512 \times 512 \time 180 \sim 300$ 2D images sliced by $0.496 mm$. Our dataset does not include small aneurysms (\textless 3.00 mm), because our objective is to segment the aneurysms requiring surgery. We calculated the size of each aneurysm based on maximum diameter, and Figure~\ref{fig:size} shows the distribution (Mean: 7.49 mm, SD: 2.72 mm; Range: 3.48–18.66 mm) of aneurysm sizes on our dataset. In terms of IA type, most of our data were saccular aneurysms, and one fusiform aneurysm was included, but no dissecting aneurysm. Another special case was that we treated two aneurysms very close together as being one. We annotated the aneurysm portions on both the entire surface models of the brain arteries and on TOF-MRA images to generate a ground truth for classification and segmentation for training neural networks.
It took a total of three experts 21 working days to perform this task.
We used five-fold cross-validation to conduct our experiments. A total of 103 sets were shuffled and divided into five subsets, of which four were used as the training data, and one was used as testing data. This study design was approved by an appropriate ethics review board.

\subsection{Evaluation metrics}
Several evaluation metrics were employed to demonstrate the performance of the models according to different tasks. Accuracy, recall, and sensitivity are typically used to evaluate performance on classification tasks. For the segmentation task, the Dice similarity coefficient (DSC) or Intersection over Union (IoU) is employed to indicate the prediction of the target region. The sensitivity of the entire input can be easily high because the target region may be tiny. Moreover, in this situation, the overall statistics may conflict with the part-wise statistics due to Simpson's paradox.

\subsection{Implementation details}
The experiments were performed on a PC with a GeForce RTX 2080Ti. During data preprocessing, the normal vector of each point was estimated using the original surface model. We also recorded the point index of the entire model as a global ID on the sampled fragments for each point to improve the efficiency of voting. We set the sampling thresholds as $\alpha = 15, \beta = 1.5*\alpha$, and the samplings in which the number of points was less than $500$ were removed. We automatically generated $7192$ vessel fragments from the 103 entire models, and $392$ fragments contained aneurysms. 

The training hyper-parameters were set as follows. For the classification network, the number of sampling points for each fragment was $1024$. The weights of the loss function were determined according to the number of fragments. We trained the network using $251$ epochs and a batch size of $8$. The classification results were predicted by setting a discrimination threshold of $0.23$. For the segmentation network, the number of sampling points was $2048$. We trained the network for up to $401$ epochs, with a batch size of $12$. For each network, we used the Adam optimizer with a learning rate of $10^{-3}$.

\subsection{Results}
\label{sec:results}

\begin{table}[t]
\caption{The sensitives of aneurysm fragments and whole IAs ($\%$).}
\label{tab:sensitive}
\centering
\begin{tabular}{c|c|c|c|c|c}
\hline
Fold & 0 & 1 & 2 & 3 & 4 \\
\hline
Fragments & 73.63 & 81.08 & 79.49 & 86.11 & 80.77 \\
Aneurysms & 95.24 & 100.00 & 100.00 & 95.00 & 85.71 \\
\hline
\end{tabular}
\end{table}

\begin{figure}[t]
\centering
\includegraphics[width=0.48\linewidth]{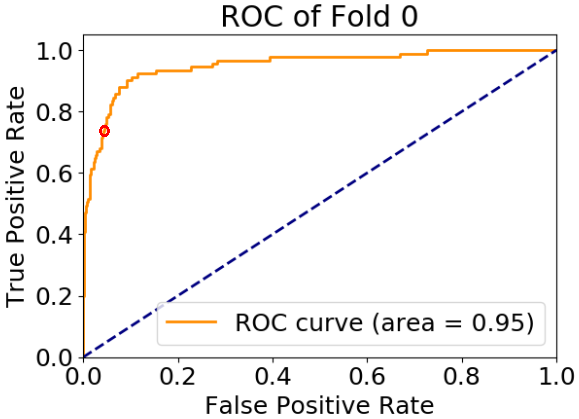}
\includegraphics[width=0.48\linewidth]{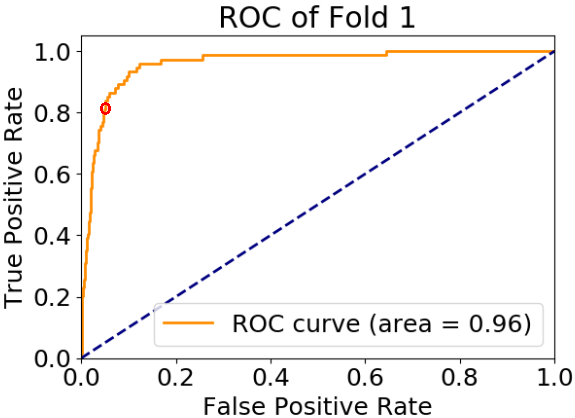}\\
\includegraphics[width=0.48\linewidth]{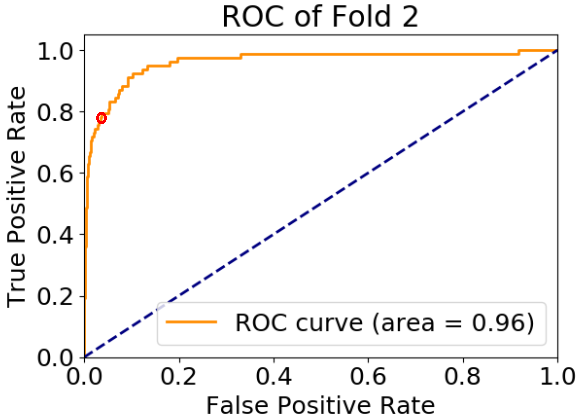}
\includegraphics[width=0.48\linewidth]{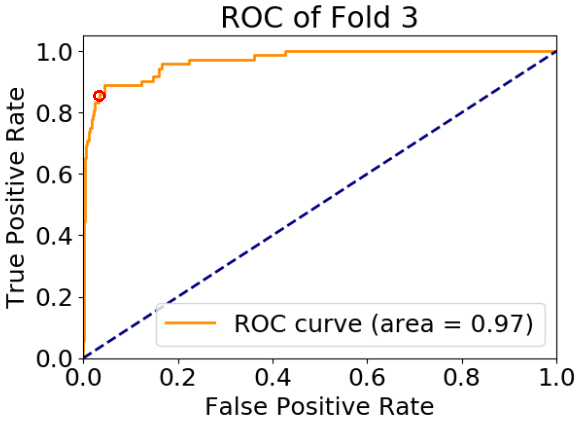}\\
\includegraphics[width=0.48\linewidth]{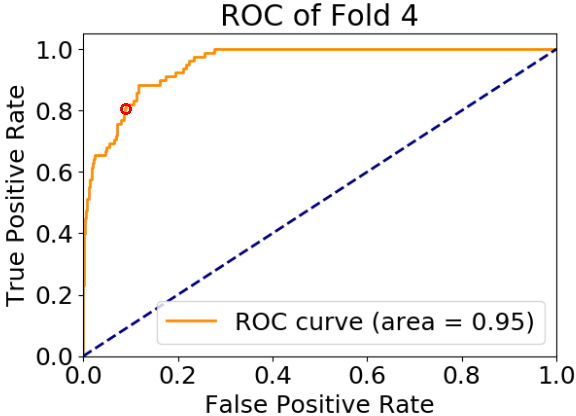}
\includegraphics[width=0.44\linewidth]{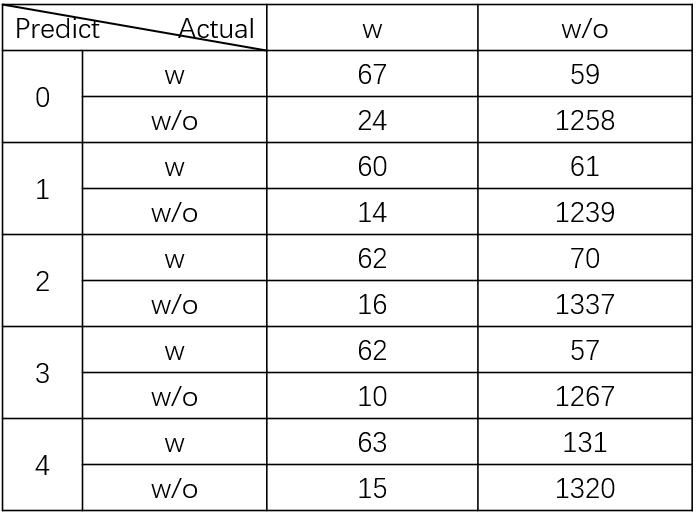}
\caption{ROC curves and confusion matrices of five trained classification networks.} 
\label{fig:roc}
\end{figure}

\begin{figure}[t]
\centering
\includegraphics[width=0.13\textwidth]{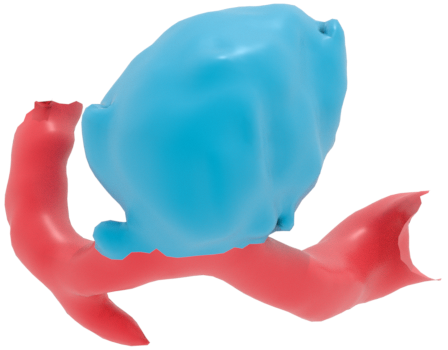}
\includegraphics[width=0.13\textwidth]{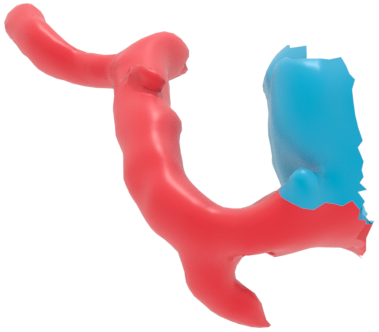}
\includegraphics[width=0.13\textwidth]{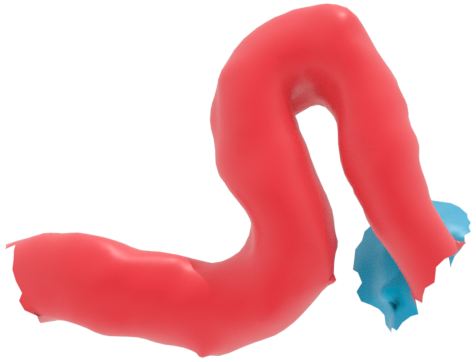}\\
A \hspace{70pt} B \hspace{60pt} C\\
\includegraphics[width=0.04\textwidth]{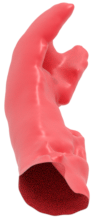}
\includegraphics[width=0.115\textwidth]{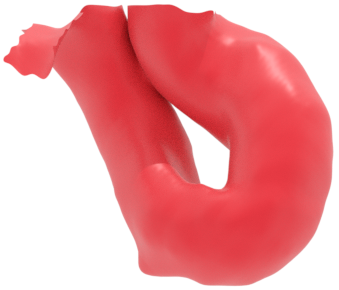}
\includegraphics[width=0.115\textwidth]{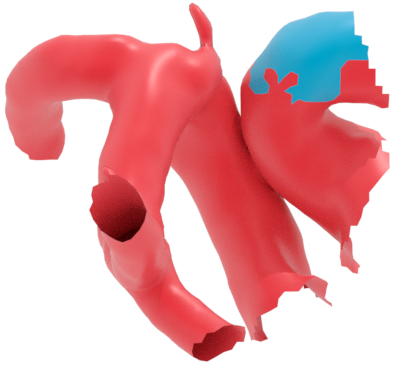}
\includegraphics[width=0.115\textwidth]{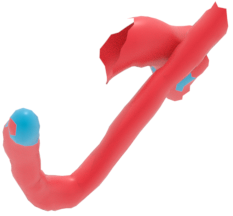}\\
% \\ A \hspace{80pt} B \hspace{70pt} C \hspace{40pt} D \hspace{50pt} E \hspace{60pt} F \hspace{60pt} G
D \hspace{40pt} E \hspace{40pt} F \hspace{50pt} G \hspace{25pt}
\caption{Fragment examples. Fragment A had a complete aneurysm. Fragments B and C partly overlap with A, but only include a part of the aneurysm. Fragments D, E, F, and G were without aneurysms, but were misclassified. The original data of D is noisy, and E, F, and G have a shape that is very similar to a part of the aneurysm.}
  \label{fig:fragments-result}
\end{figure}

%We show the experimental results for the trained networks with the best performance. 
The receiver operating characteristic (ROC) curve and confusion matrix of each classification network are shown in Figure~\ref{fig:roc}.
% We first show (About) the results of our classification network. The receiver operating characteristic (ROC) curve and confusion matrix of each classification network is shown in Figure~\ref{fig:roc}. 
It can be observed that all areas of the ROC curves are higher than 0.95, demonstrating that the trained classification networks are generalized. The sensitivities to the aneurysm class of the five networks were $73.63\%$, $81.08\%$, $79.49\%$, $86.11\%$, and $80.77\%$, respectively. This shows that we can precisely detect fragments with IA portions.
By analyzing the confusion matrices, we observed that only a few fragments with aneurysms were misclassified because they had tiny aneurysms or contained only a small part of the aneurysm. However, a $100\%$ sensitivity is not necessary for our classification network because the sampled fragments overlap, as shown in Figure~\ref{fig:fragments-result}. According to our sampling algorithm, the $80\%$ sensitivity of the classifier does not mean that 1/5 of the aneurysms were already missed before the segmentation step. In this experiment, only 5 out of 114 IAs were missing. The real sensitivity to IAs is satisfactory, as shown in Table~\ref{tab:sensitive}. However, some fragments without IA were not classified correctly because the original data contained significant noise and the fragments had a shape very similar to a small part of the full aneurysms. These misclassified cases were also sometimes difficult for segmentation networks. Therefore, they did not have a significant impact on the final segmentation results. 

We added the classification step before segmentation to filter out the majority of fragments that did not contain aneurysms. This helped the segmentation network to avoid predicting false positive results on non-aneurysm regions, as well as, improving the balance between fragments with and without aneurysms, leading to better final segmentation accuracy. The benefit of the classification process is also shown in Section~\ref{subsec:seg_only}, in which we compare the proposed two-step pipeline and a segmentation-only pipeline.
% The classification step is valuable, the details of the comparison experiment between our proposed two-step design and without classification  step are described in Section~\ref{subsec:seg_only}. By filtering out the majority of fragments that do not contain aneurysms, we cannot only save a lot of training time but also improve the accuracy of the final segmentation result.}

\begin{figure}[th]
% \centering
\begin{center}
\includegraphics[width=0.49\linewidth]{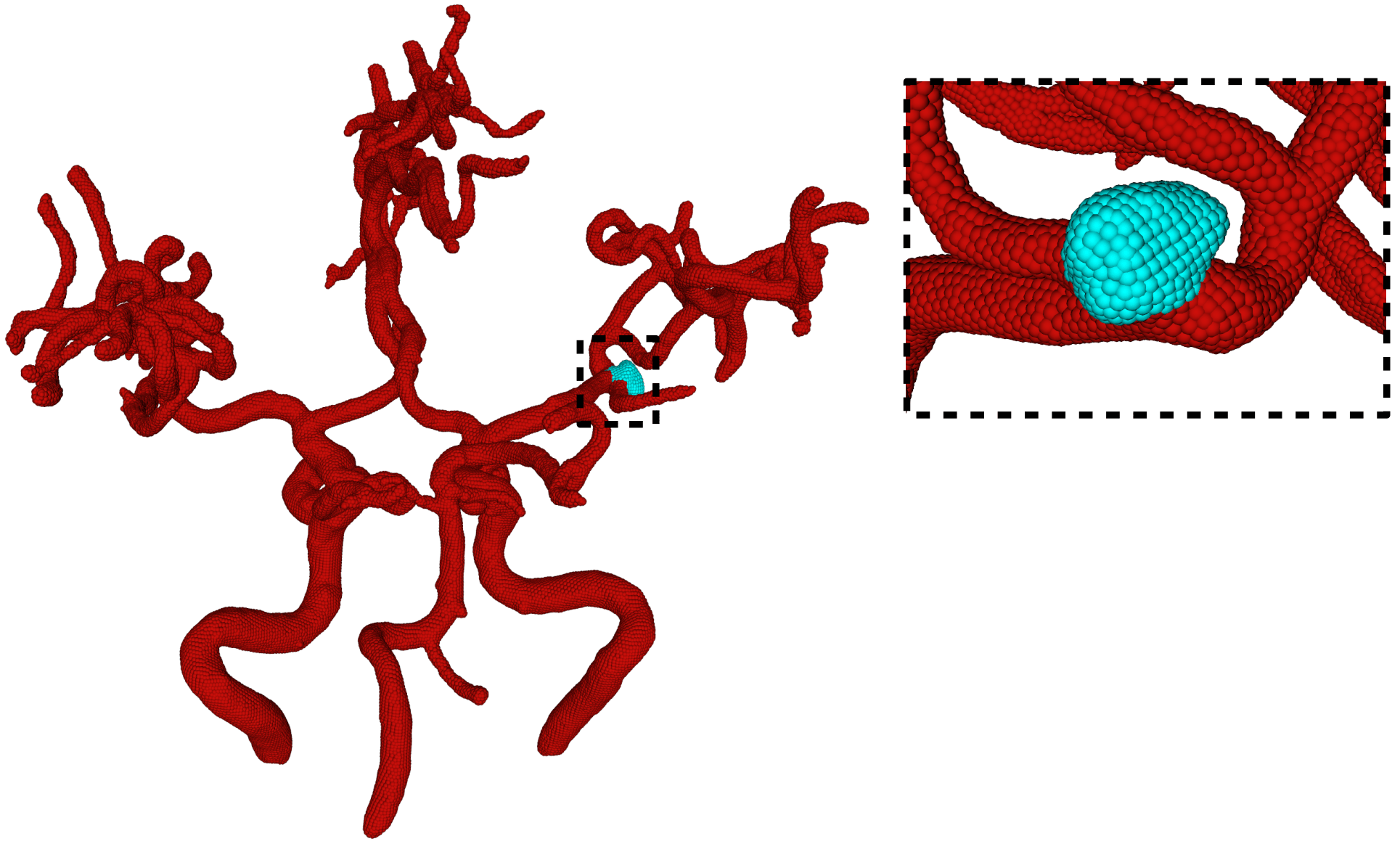}~~~
\includegraphics[width=0.49\linewidth]{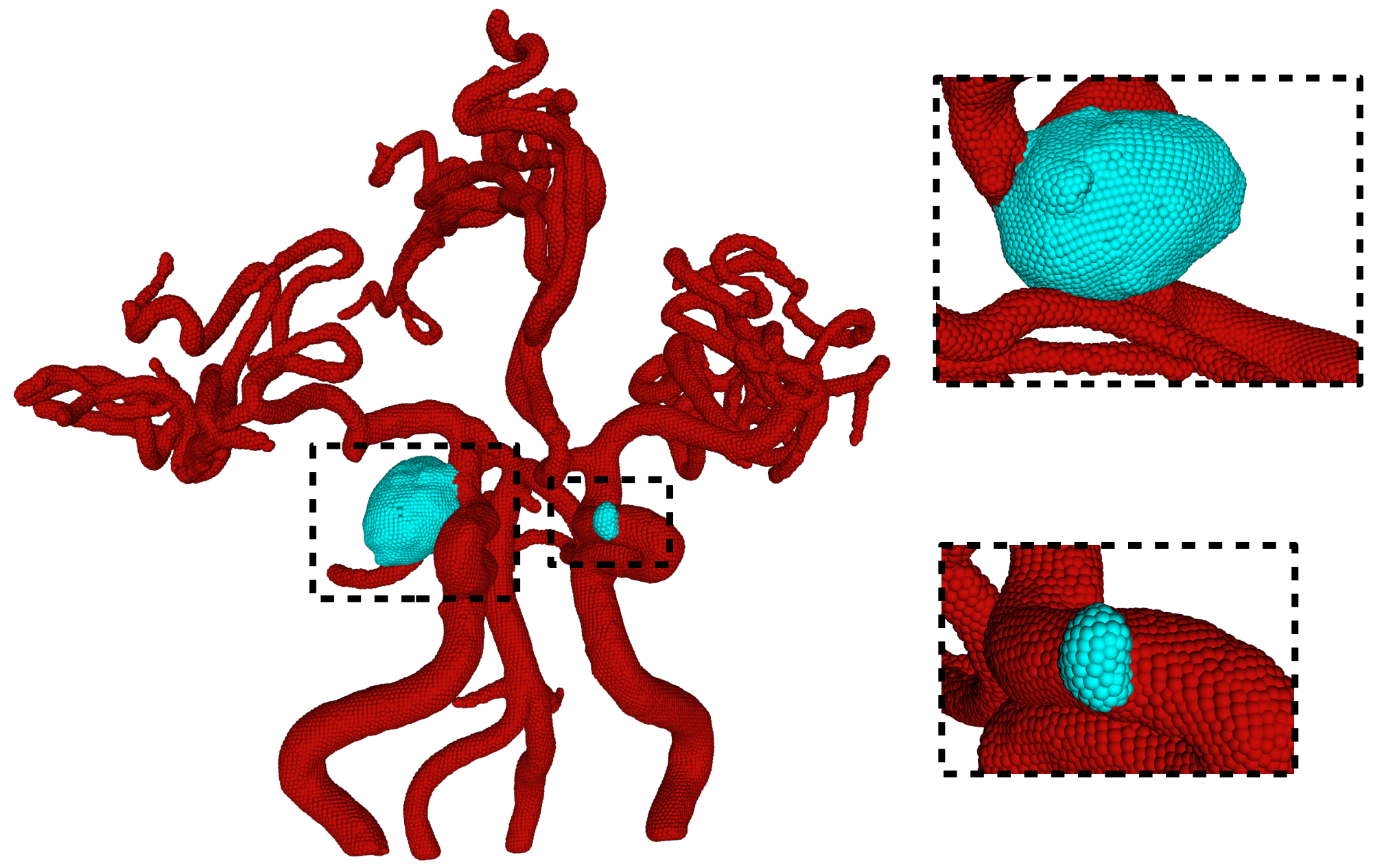}
\end{center}
\justify{Left: Saccular aneurysms. The proposed pipeline obtained a perfect segmentation. Right: Saccular aneurysm. The aneurysm was segmented clearly without the impact of close blood vessels as shown in the top enlarged figure. A potential aneurysm was also predicted, which was not annotated by experts, as shown in the bottom enlarged figure.}
\begin{center}
\includegraphics[width=0.49\linewidth]{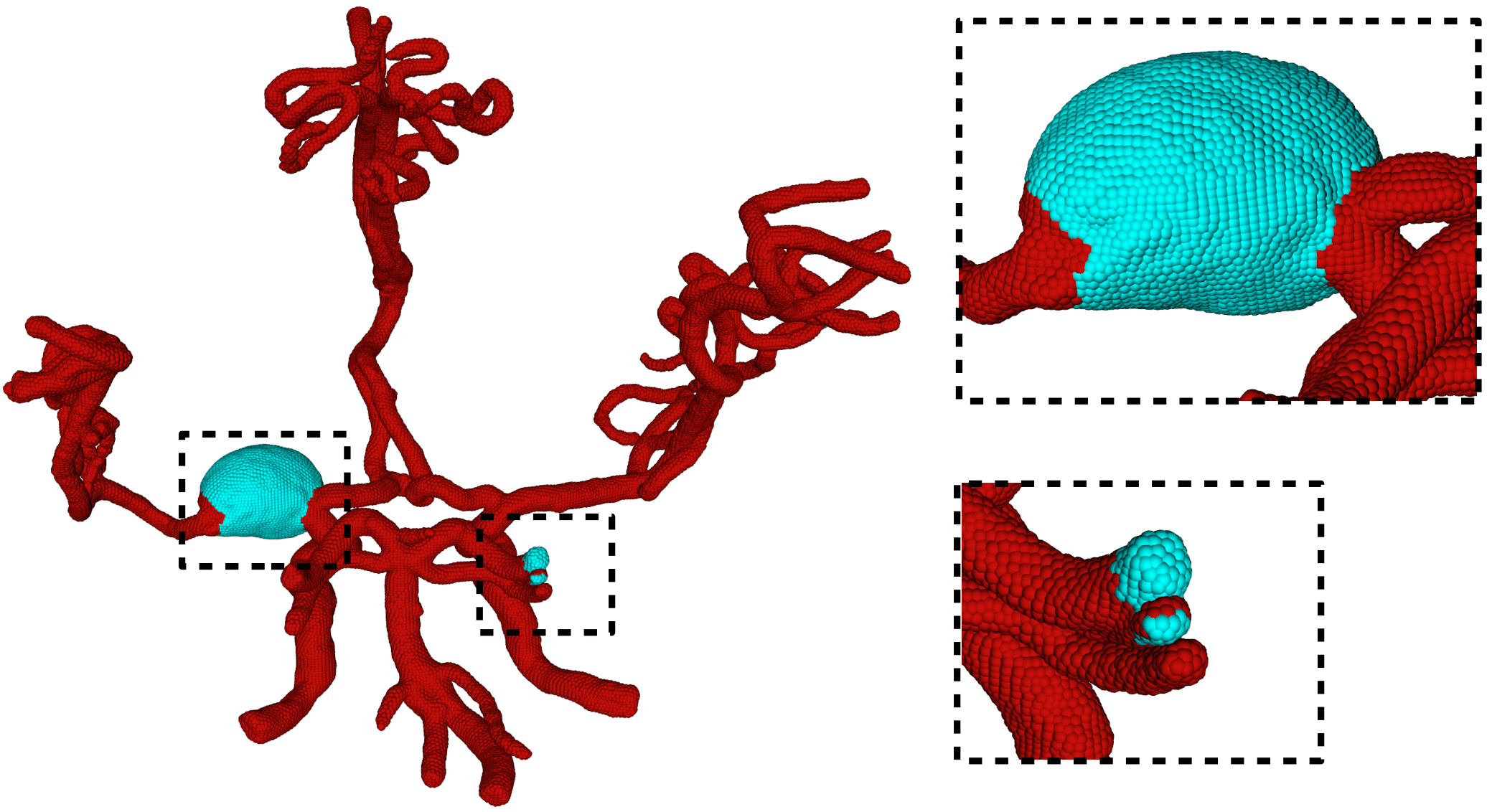}~~~
\includegraphics[width=0.49\linewidth]{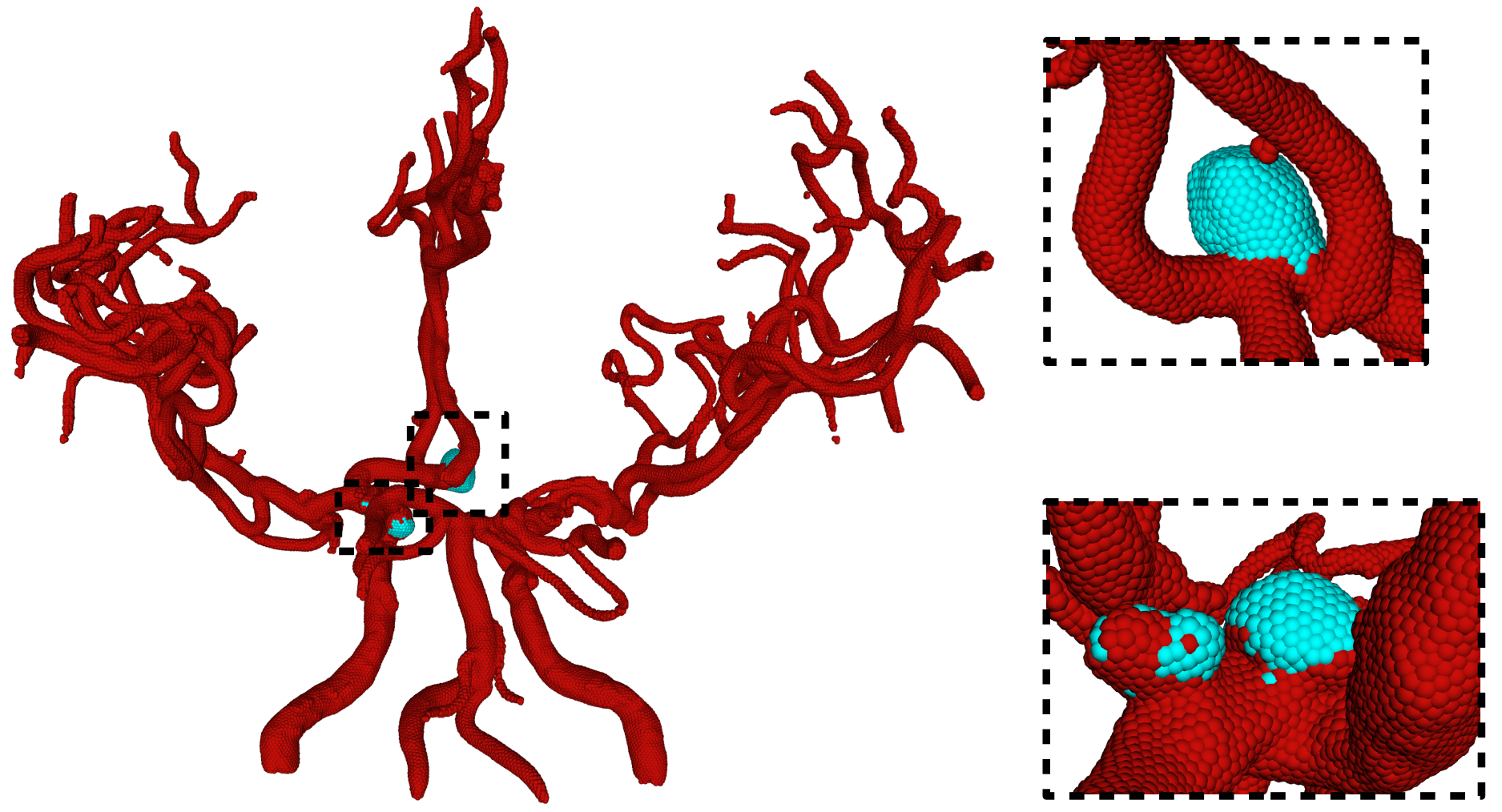}
\end{center}
\justify{Left: Fusiform aneurysms. The aneurysm was well segmented, as shown in the top enlarged figure. The ends of normal blood vessels were segmented incorrectly into aneurysms, as shown in the bottom figure. Right: Double saccular aneurysm; multiple aneurysms in one case. Two close aneurysms were annotated as one, and our segmentation did not achieve the same result as Yang et al.~\cite{yang2020intra} because our training data included much more complex shapes.}
\caption{Examples of final segmentation results (rendering in points); enlarged figures may be captured from a different viewpoint to show the aneurysm shape.}
\label{fig:shape}
\end{figure}

Four examples of the final segmentation results are presented in Figure~\ref{fig:shape}. In this figure, we show the entire 3D surface models and the enlarged important parts marked by black dotted boxes. The segmented aneurysms are colored in cyan, and other normal blood vessels are colored in red. We can see that our proposed pipeline obtained satisfactory segmentation results for various shapes and sizes of saccular aneurysms. 
%\textcolor{red}{Particularly, the close blood vessels did not affect the segmentation results according to the feature that the points-based network could learn manifolds.} 
We also found that unannotated potential aneurysms could also be segmented. However, a small amount of normal vessel ends were segmented as aneurysms because their shape were extremely similar to IAs. In addition, our networks predicted a suitable segmentation result for the fusiform aneurysm, even though they were trained only on saccular aneurysms. This demonstrates the excellent generalization ability of point-based deep learning models. Furthermore, our proposed pipeline obtained superb segmentation of multiple aneurysms in one case. A more detailed statistical analysis of our final segmentation results is provided in the comparison experiments~\ref{sec:comparison}.

\subsection{Comparison experiments}
\label{sec:comparison}

\subsubsection{3D U-Net}
We first applied the original 3D U-Net to our data; however, the network cannot predict any segmentation result because the aneurysms were too small compared to the entire image. This experiment demonstrates the difficulty of this segmentation task.

\subsubsection{Segmentation only}
\label{subsec:seg_only}

To indicate the necessity of the two-step design, we performed an ablation study by removing the classification step from our proposed pipeline. That is, we fed all fragments to the pre-trained segmentation network to segment the aneurysm regions in the fragments. We compared the final results of this segmentation-only pipeline with those of the two-step pipeline on the entire artery surface models as shown in Table~\ref{tab:comparison_two}. The comparison results demonstrate that the classification step greatly reduces noise and improves the performance of the final segmentation result.

\begin{table}[t]
\caption{Comparison of segmentation results between segmentation only and our two-step design on surface DSCs ($\%$).}
\label{tab:comparison_two}
\centering
\begin{tabular}{c|c|c|c|c}
\hline
& \multicolumn{2}{c|}{Segmentation only} & \multicolumn{2}{c}{Two-steps} \\
\hline
& Mean & STD & Mean & STD \\
\hline
Overall & 31.43 & 16.92 & \textbf{74.74} & 26.47 \\ 
\hline
Fold 0 & 36.10 & 15.63 & \textbf{76.73} & 25.83 \\
Fold 1 & 38.21 & 19.28 & \textbf{80.18} & 17.75 \\
Fold 2 & 33.74 & 18.81 & \textbf{80.66} & 20.33 \\
Fold 3 & 26.67 & 11.78 & \textbf{73.78} & 25.67 \\
Fold 4 & 22.52 & 14.00 & \textbf{62.54} & 36.62 \\
\hline
\end{tabular}
\end{table}

\subsubsection{DeepMedic}

We also applied the method described in~\cite{sichtermann2019deep} to our dataset for comparison. For preprocessing approaches $A, B, C$, and $D$ were applied in their study. $A$ has only been applied as a necessary step in DeepMedic, while $B, C$, and $D$, were used as additional masks for the skull-stripping of the TOF-MRA images. $B$ generated the masks with a fixed threshold, $C$ used a manual threshold for the skull-stripping of each sample, and $D$ added N4 bias correction to the result of $C$. By analyzing the segmentation results, we found that skull-stripping could improve performance; however, there was not much difference between the results of $B$, $C$, and $D$. Therefore, we compared our method with $B$, which has the highest reproducibility. We used BET2 to obtain masks for skull-stripping using a fixed threshold of $0.2$. The input of the TOF-MRA images was resized to $256 \times 256$ by down-sampling, according to the requirements of DeepMedic.

\begin{figure*}[t]
\centering
\underline{Ours}\\
GT:~~~
\includegraphics[width=0.13\textwidth]{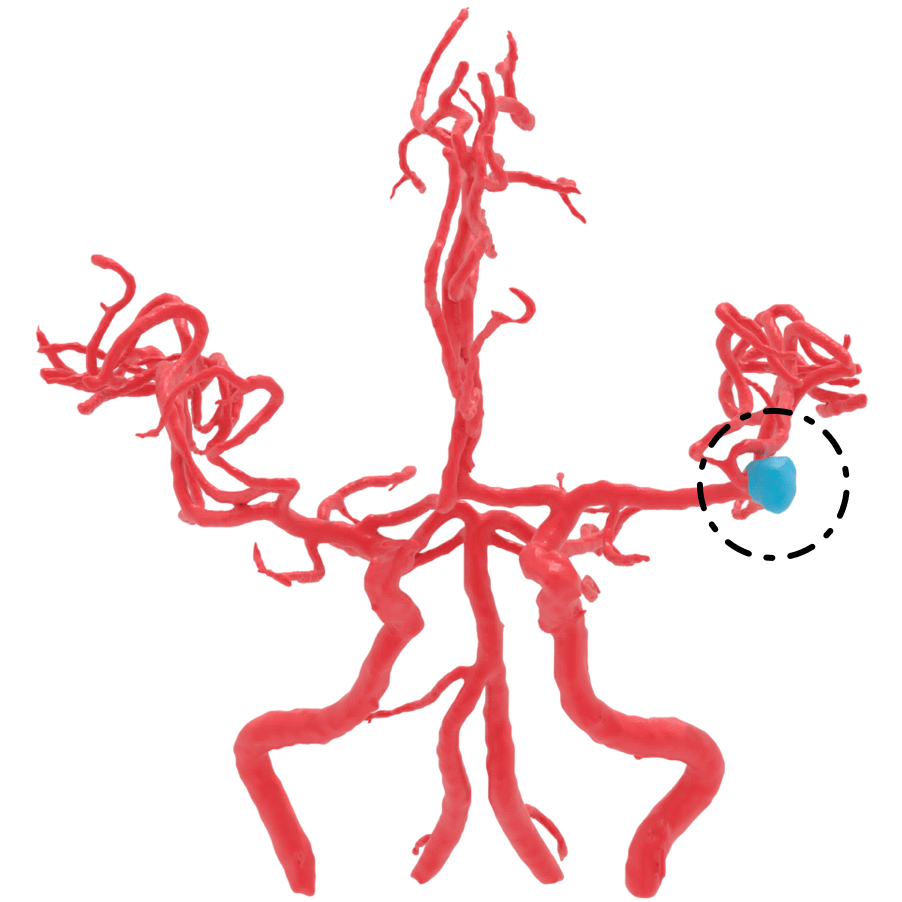}~~~
\includegraphics[width=0.13\textwidth]{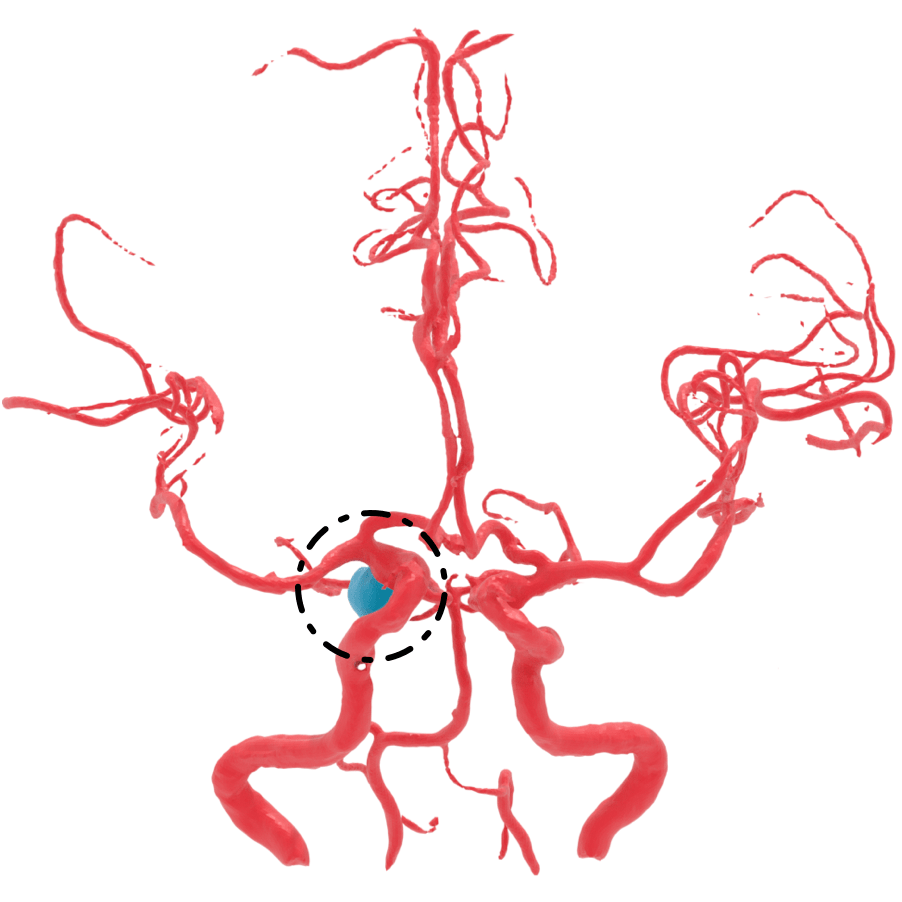}~~~
\includegraphics[width=0.13\textwidth]{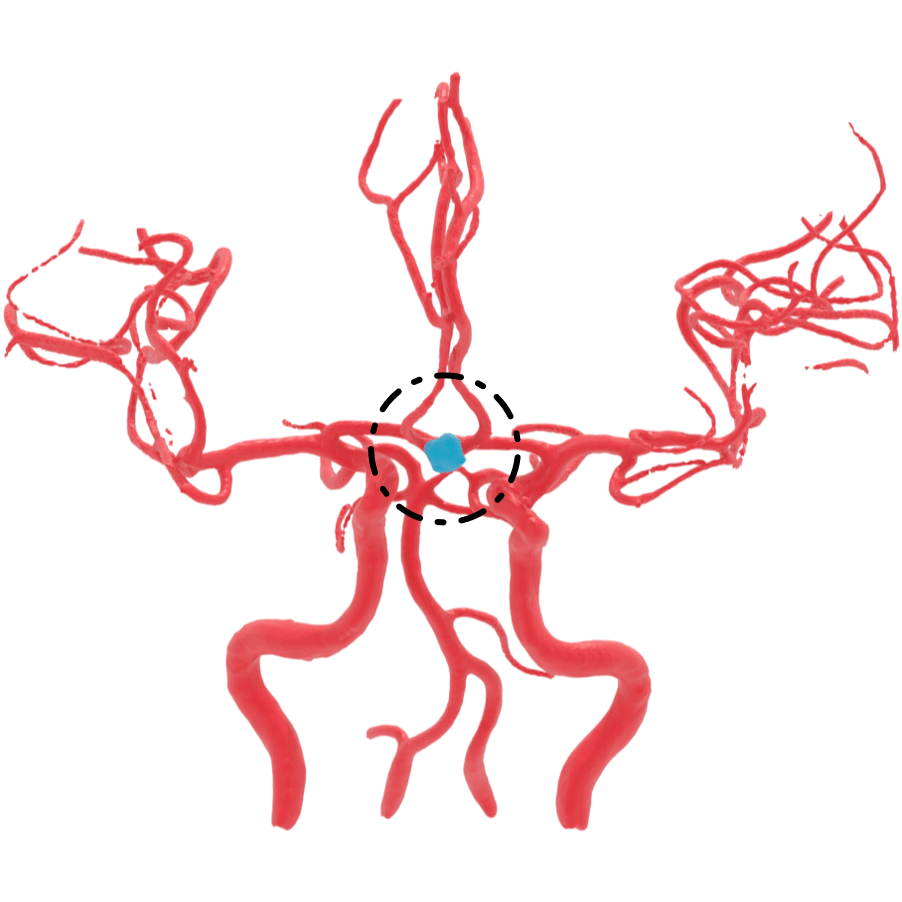}~~~
\includegraphics[width=0.13\textwidth]{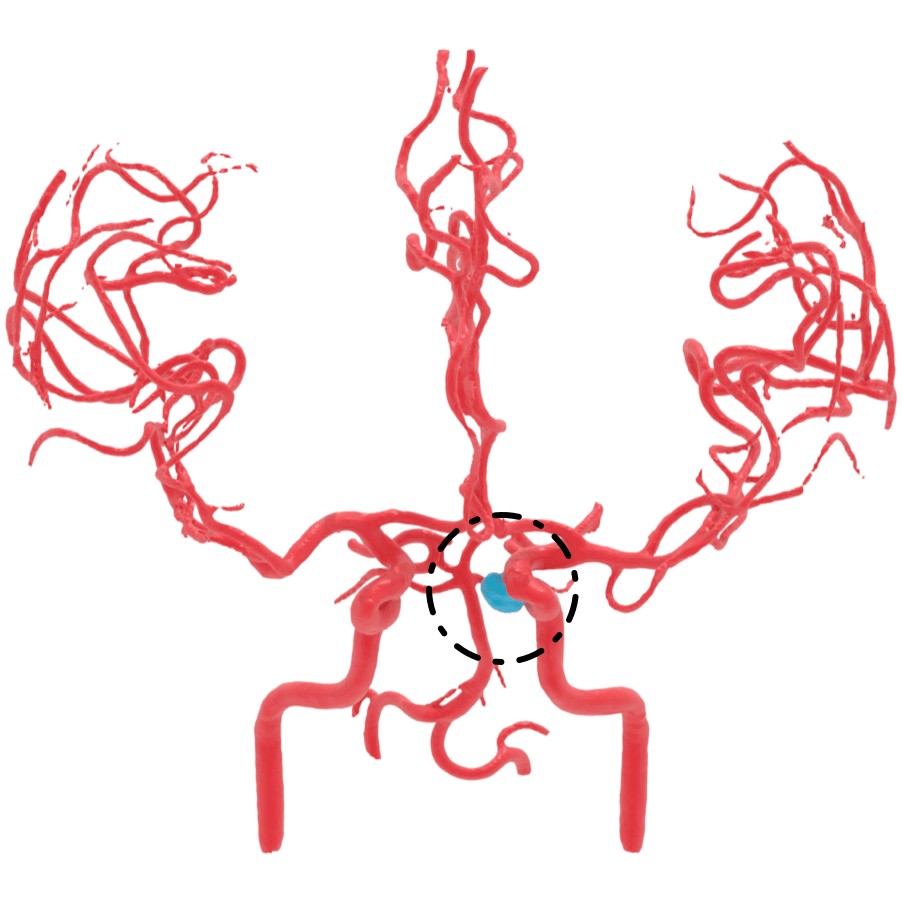}~~~
\includegraphics[width=0.13\textwidth]{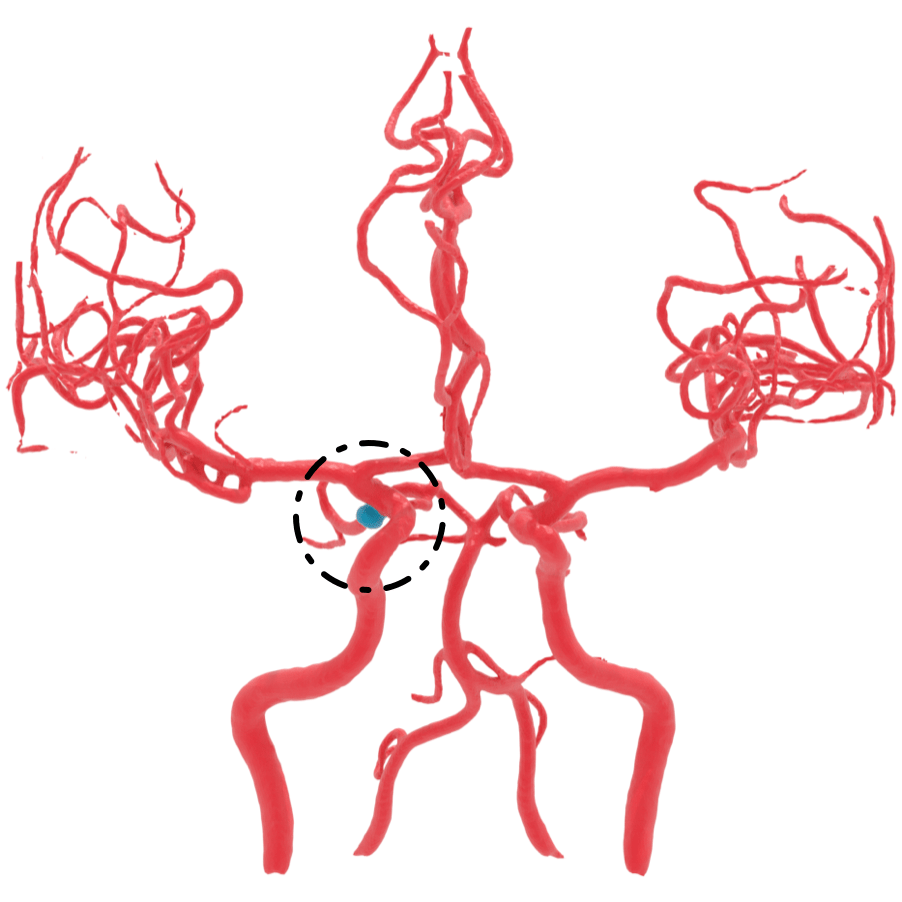}~~~
\includegraphics[width=0.13\textwidth]{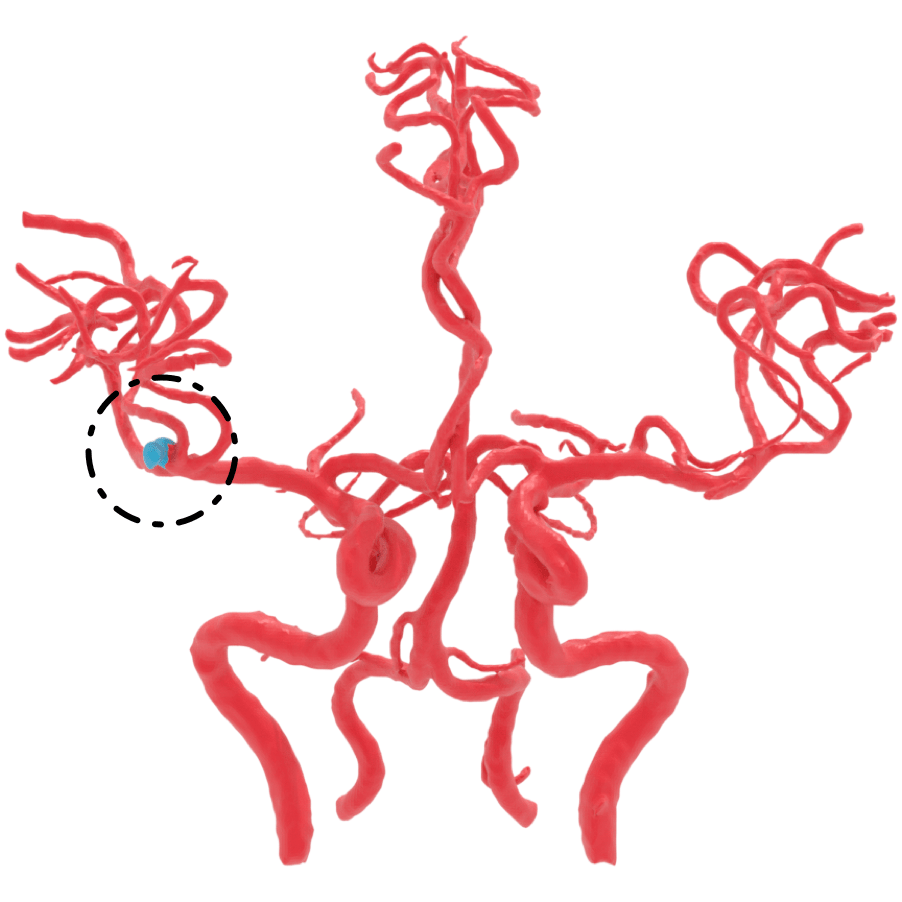}\\
Pred:~
\includegraphics[width=0.13\textwidth]{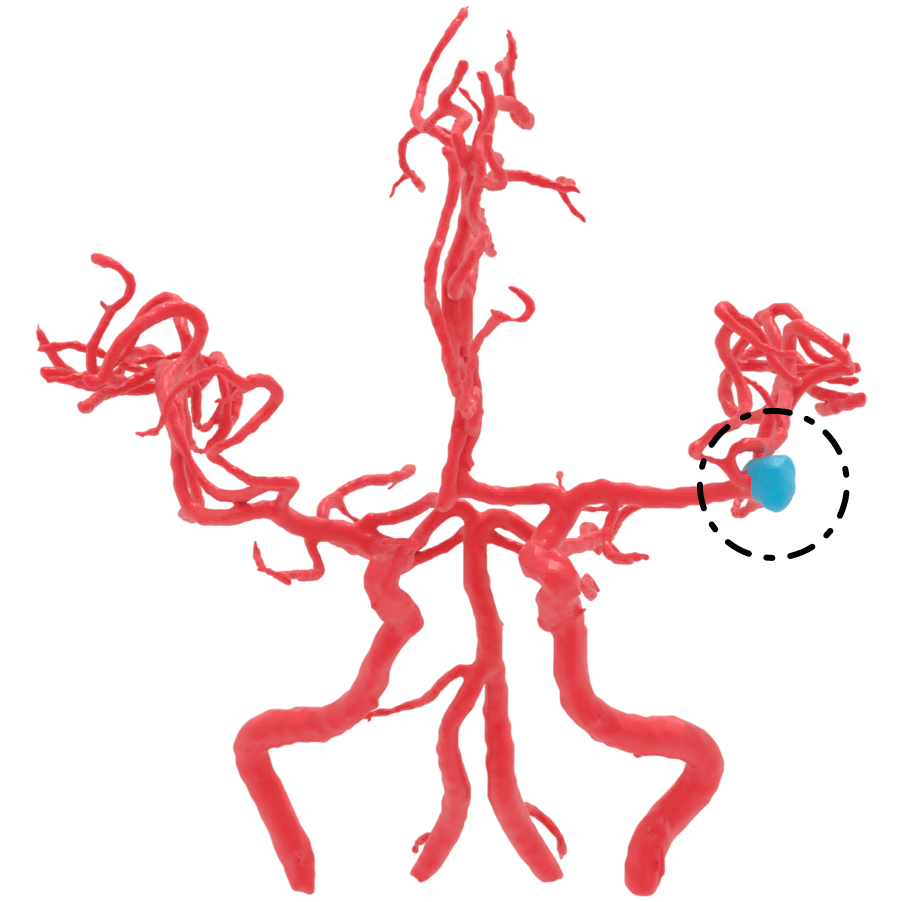}~~~
\includegraphics[width=0.13\textwidth]{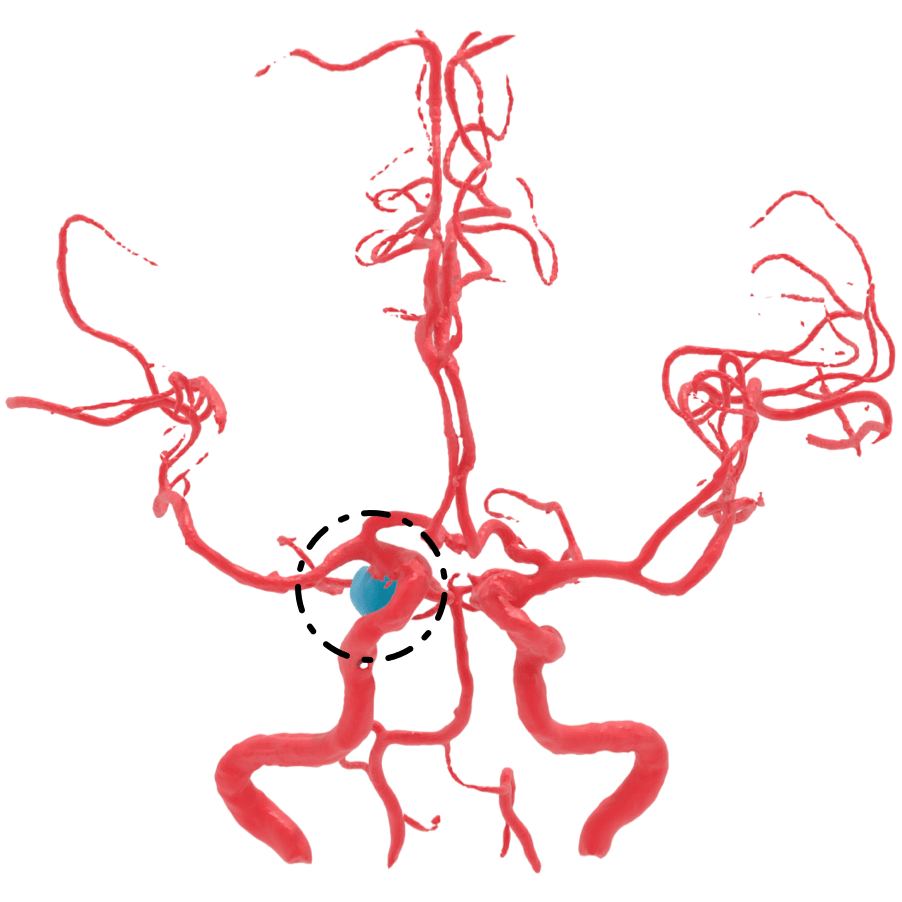}~~~
\includegraphics[width=0.13\textwidth]{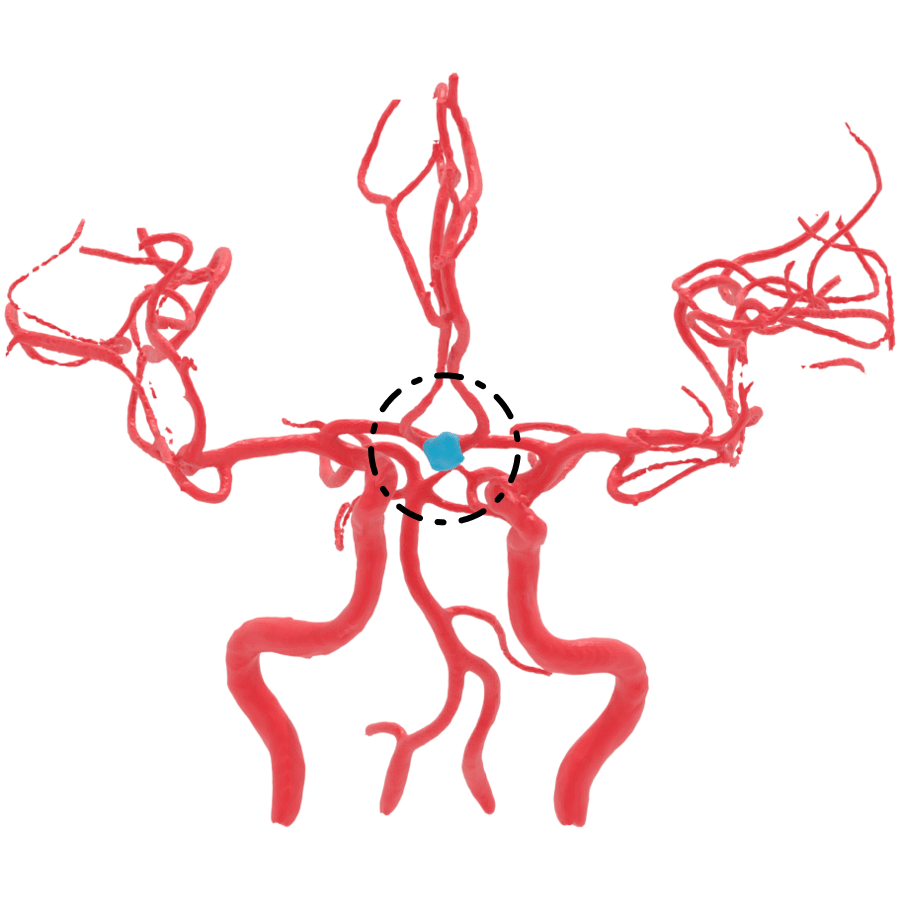}~~~
\includegraphics[width=0.13\textwidth]{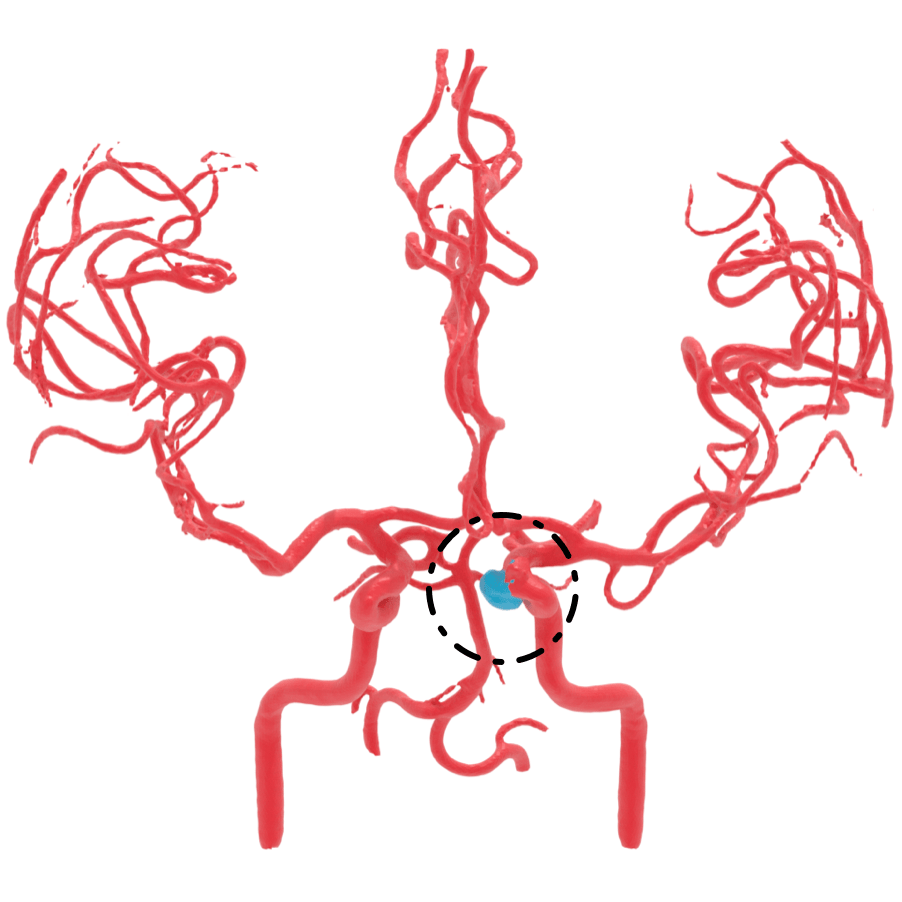}~~~
\includegraphics[width=0.13\textwidth]{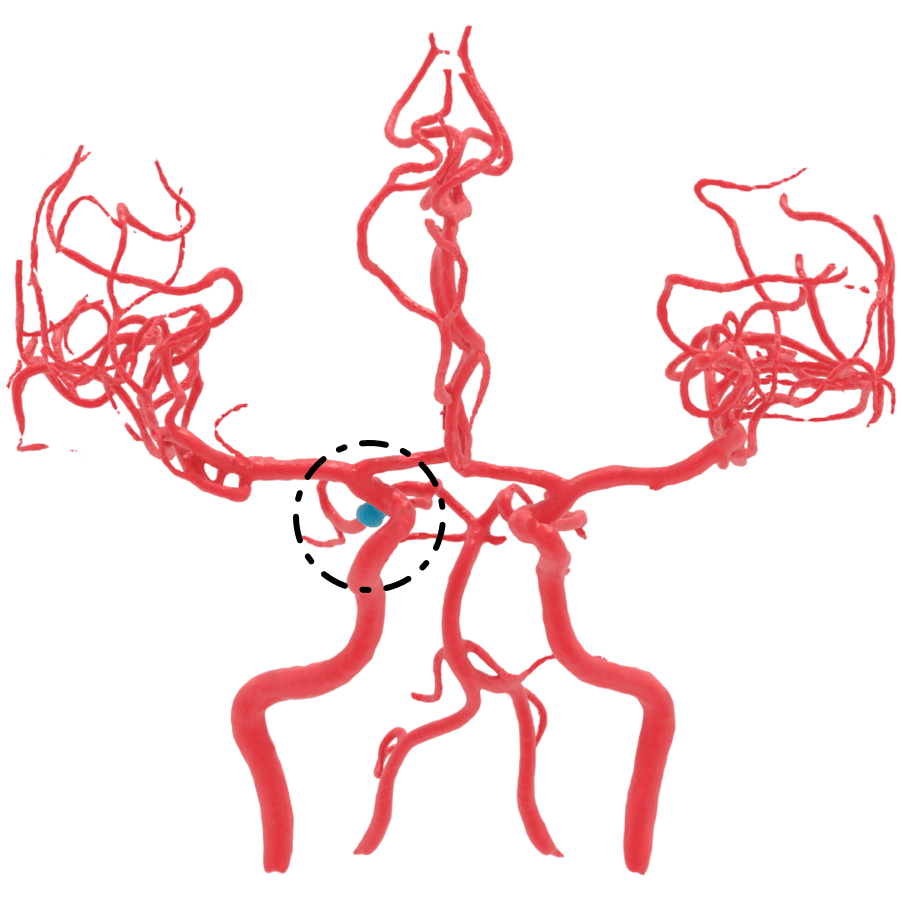}~~~
\includegraphics[width=0.13\textwidth]{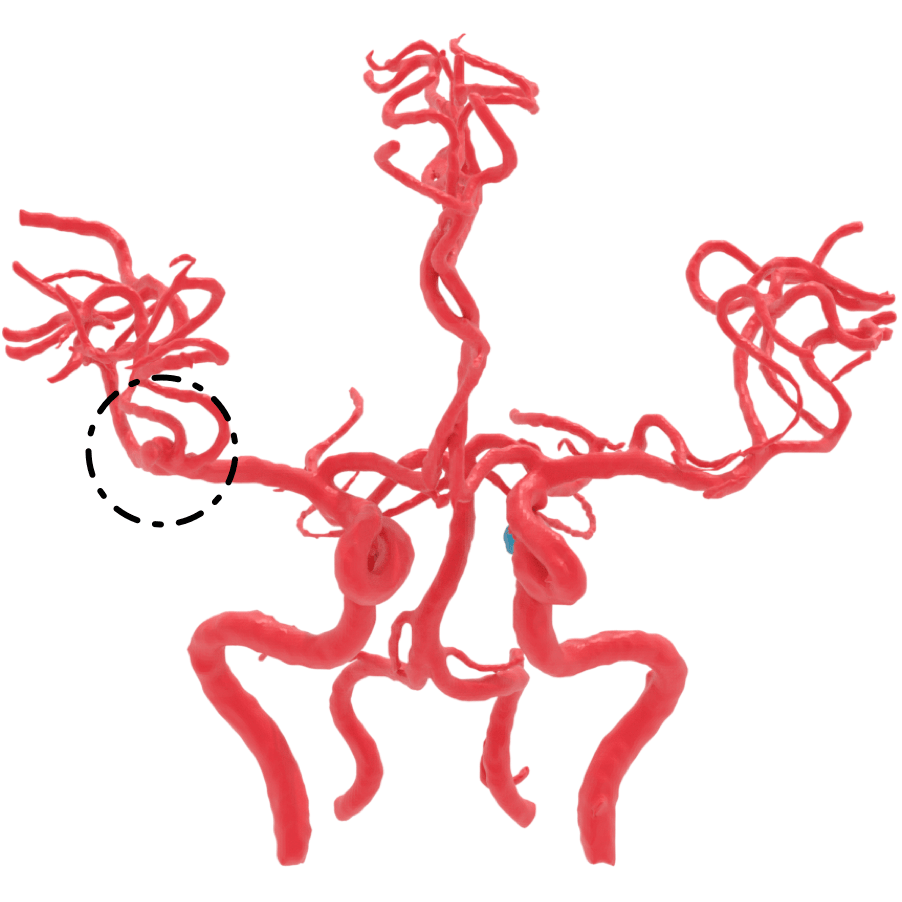}\\
GT:~~~
\begin{minipage}[t]{0.13\textwidth}
  \centering
  \includegraphics[width=0.7\textwidth]{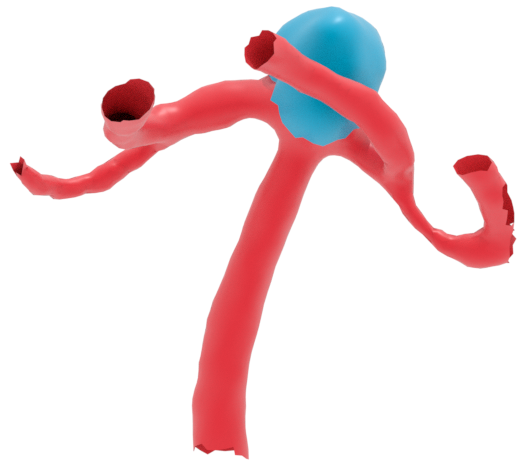}
\end{minipage}~~~
\begin{minipage}[t]{0.13\textwidth}
  \centering
  \includegraphics[width=0.7\textwidth]{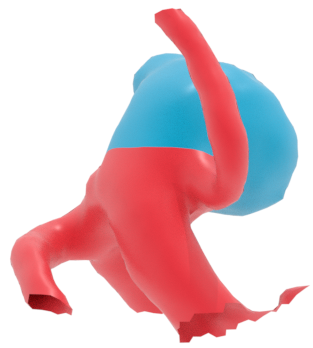}
\end{minipage}~~~
\begin{minipage}[t]{0.13\textwidth}
  \centering
  \includegraphics[width=0.7\textwidth]{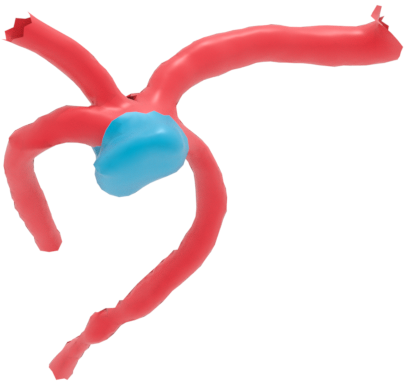}
\end{minipage}~~~
\begin{minipage}[t]{0.13\textwidth}
  \centering
  \includegraphics[width=0.7\textwidth]{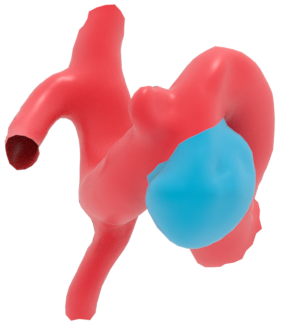}
\end{minipage}~~~
\begin{minipage}[t]{0.13\textwidth}
  \centering
  \raisebox{0.5\height}{\includegraphics[width=0.7\textwidth]{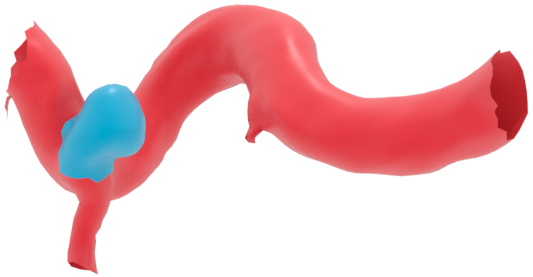}}
\end{minipage}~~~
\begin{minipage}[t]{0.13\textwidth}
  \centering
  \includegraphics[width=0.7\textwidth]{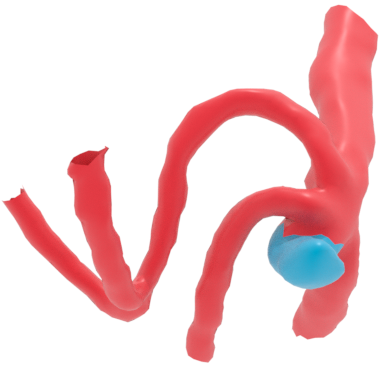}
\end{minipage}\\
Pred:~
\begin{minipage}[t]{0.13\textwidth}
  \centering
  \includegraphics[width=0.7\textwidth]{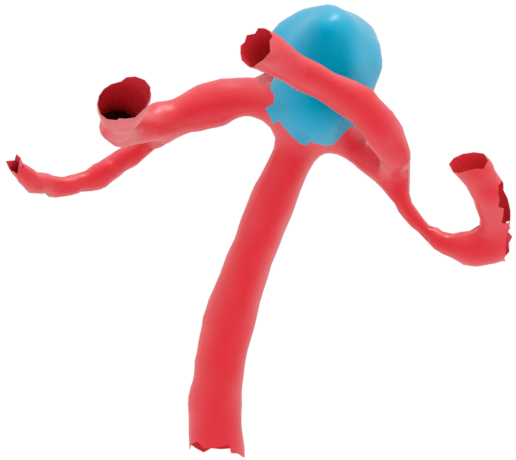}
\end{minipage}~~~
\begin{minipage}[t]{0.13\textwidth}
  \centering
  \includegraphics[width=0.7\textwidth]{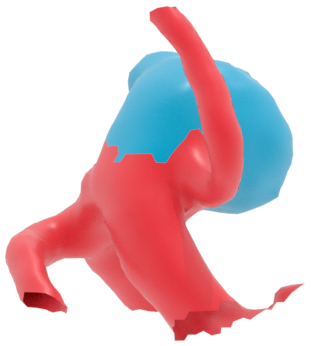}
\end{minipage}~~~
\begin{minipage}[t]{0.13\textwidth}
  \centering
  \includegraphics[width=0.7\textwidth]{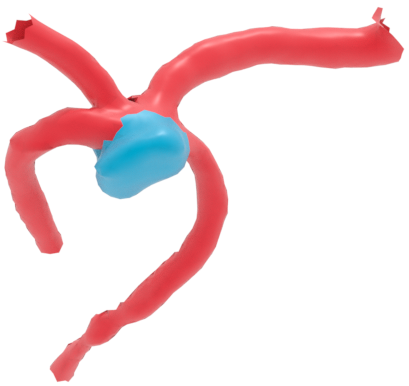}
\end{minipage}~~~
\begin{minipage}[t]{0.13\textwidth}
  \centering
  \includegraphics[width=0.7\textwidth]{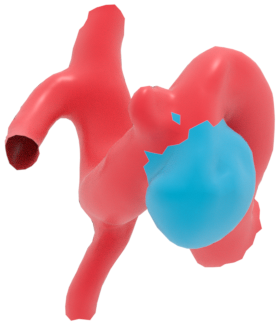}
\end{minipage}~~~
\begin{minipage}[t]{0.13\textwidth}
  \centering
  \raisebox{0.5\height}{\includegraphics[width=0.7\textwidth]{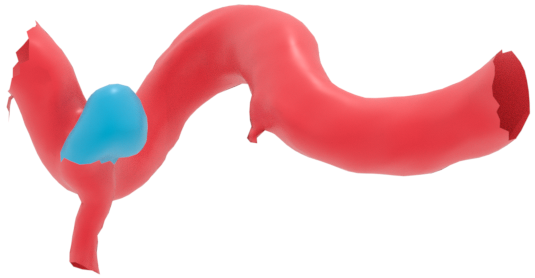}}
\end{minipage}~~~
\begin{minipage}[t]{0.13\textwidth}
  \centering
  \includegraphics[width=0.7\textwidth]{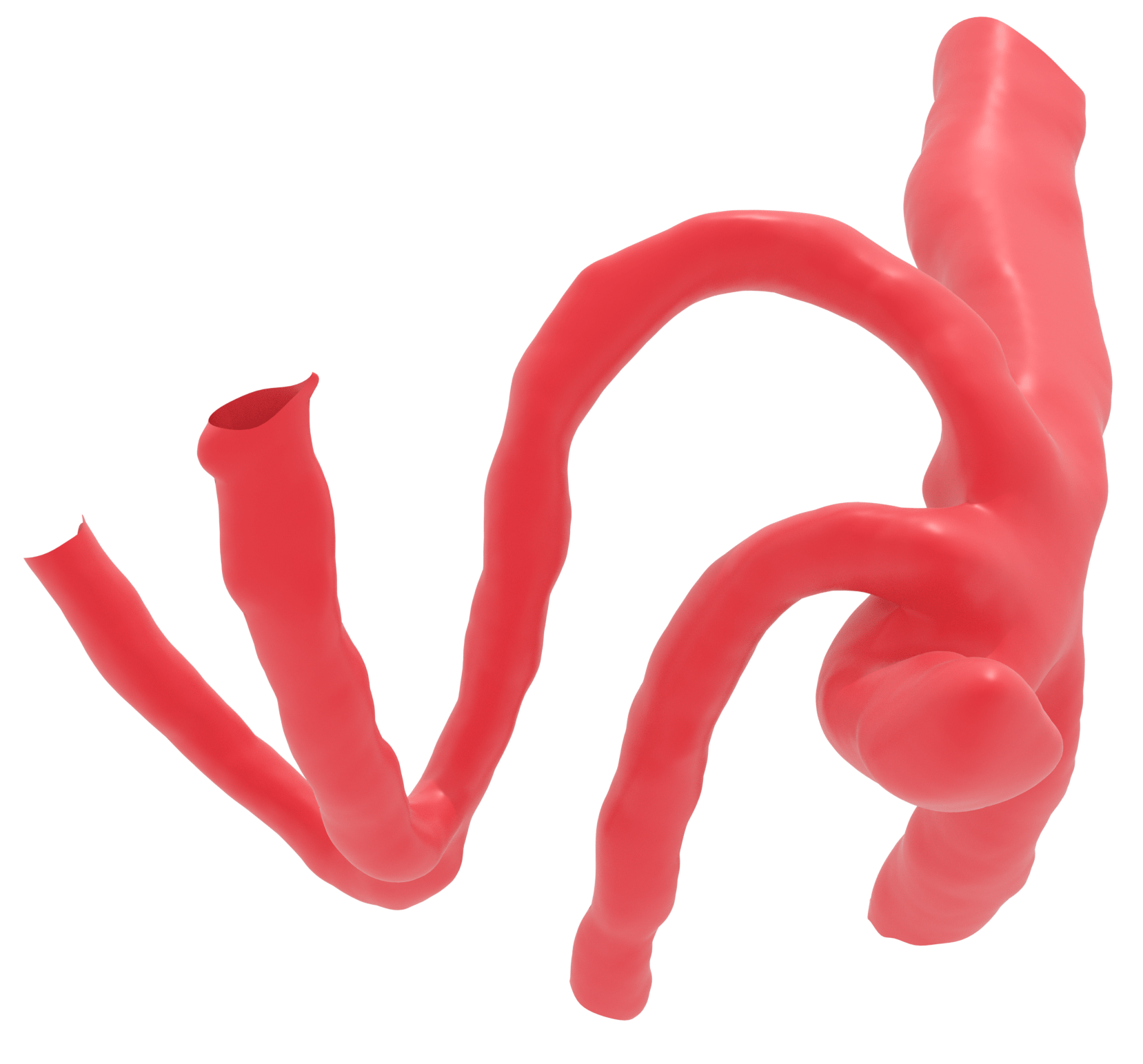}
\end{minipage}\\

\underline{DeepMedic}\\
Input:
\includegraphics[width=0.13\textwidth]{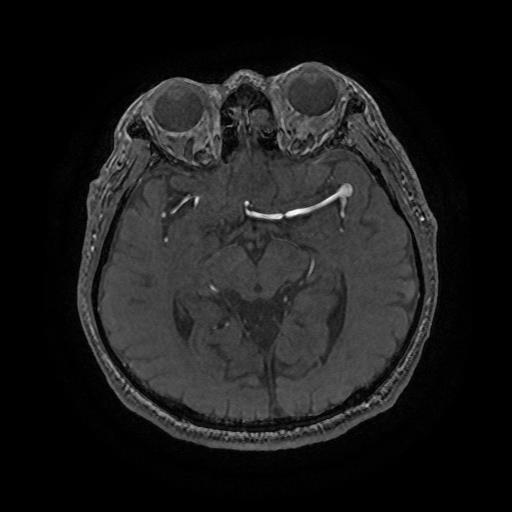}~~~
\includegraphics[width=0.13\textwidth]{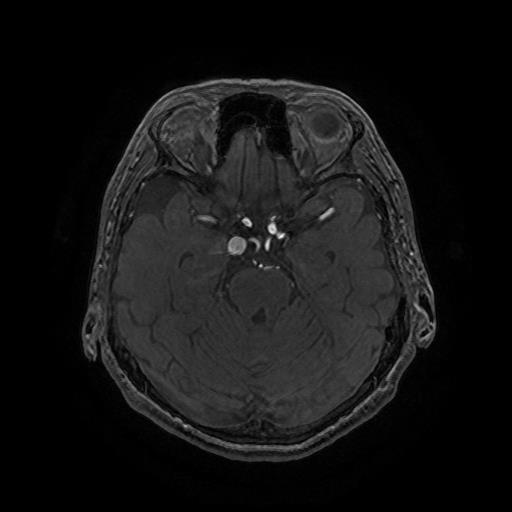}~~~
\includegraphics[width=0.13\textwidth]{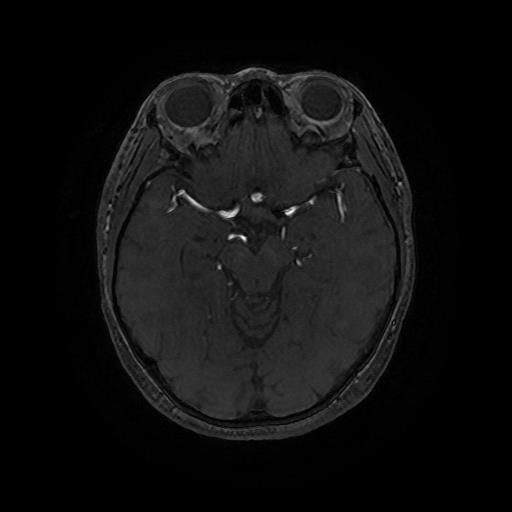}~~~
\includegraphics[width=0.13\textwidth]{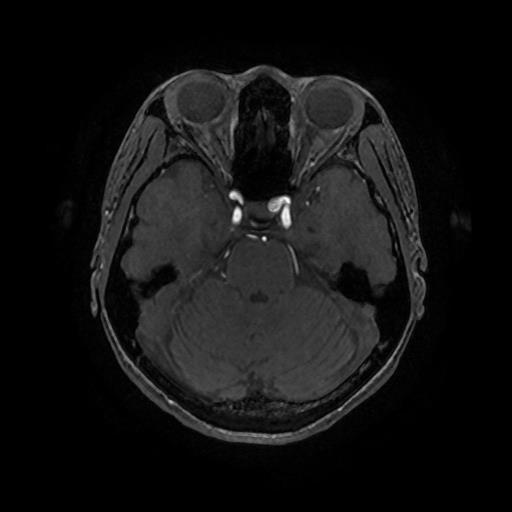}~~~
\includegraphics[width=0.13\textwidth]{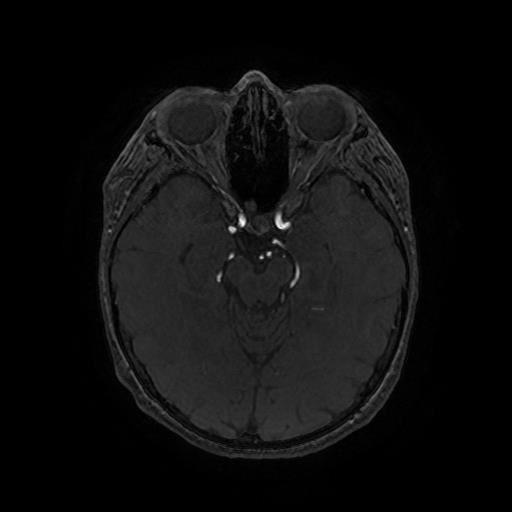}~~~
\includegraphics[width=0.13\textwidth]{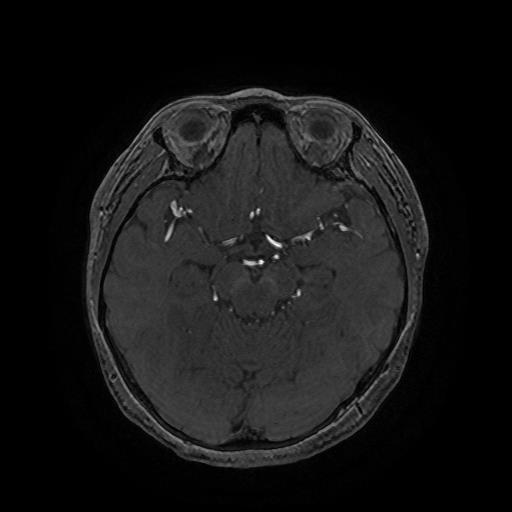}\\
\hspace{8pt}GT:~~~
\includegraphics[width=0.13\textwidth]{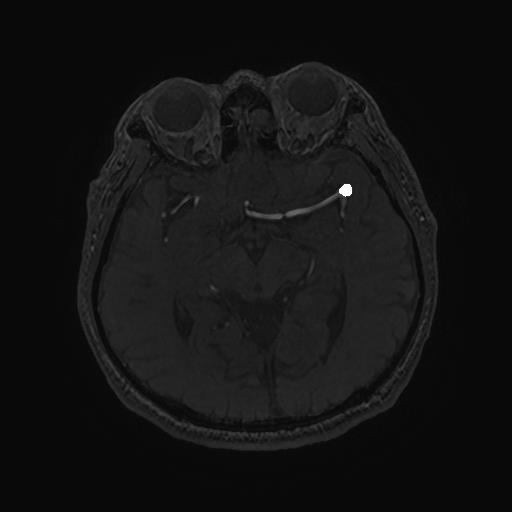}~~~
\includegraphics[width=0.13\textwidth]{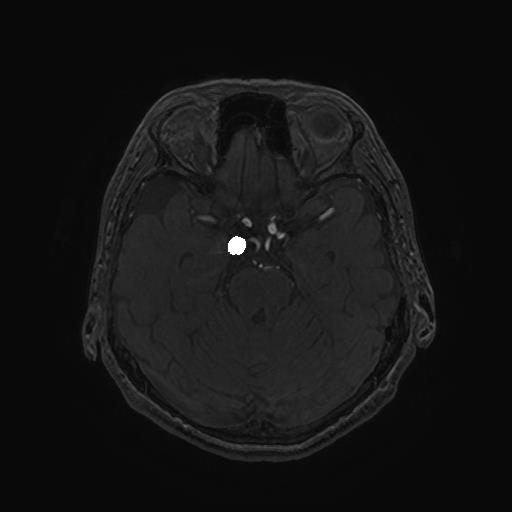}~~~
\includegraphics[width=0.13\textwidth]{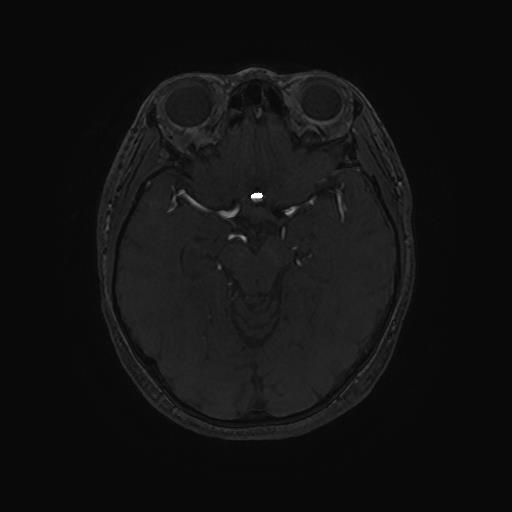}~~~
\includegraphics[width=0.13\textwidth]{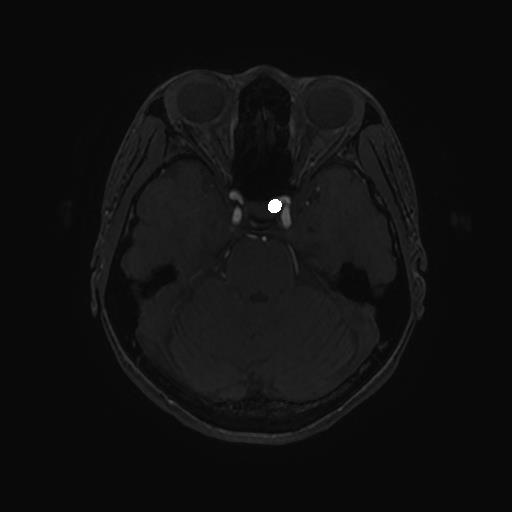}~~~
\includegraphics[width=0.13\textwidth]{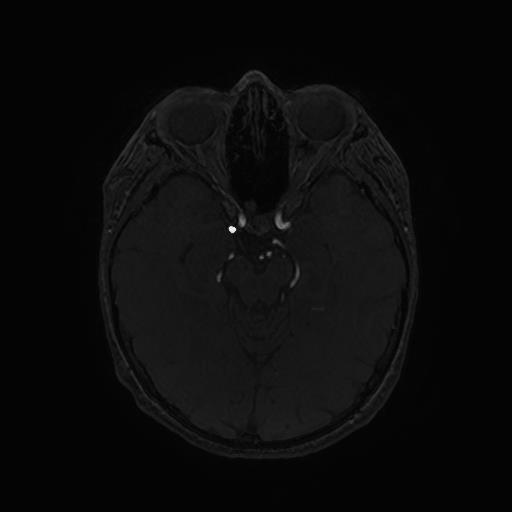}~~~
\includegraphics[width=0.13\textwidth]{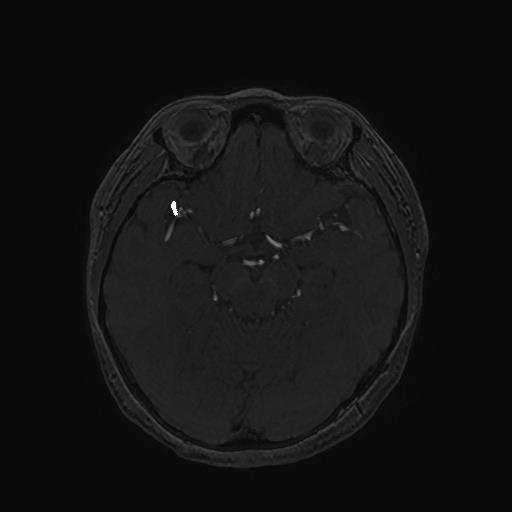}\\
\hspace{2pt}Pred:~
\includegraphics[width=0.13\textwidth]{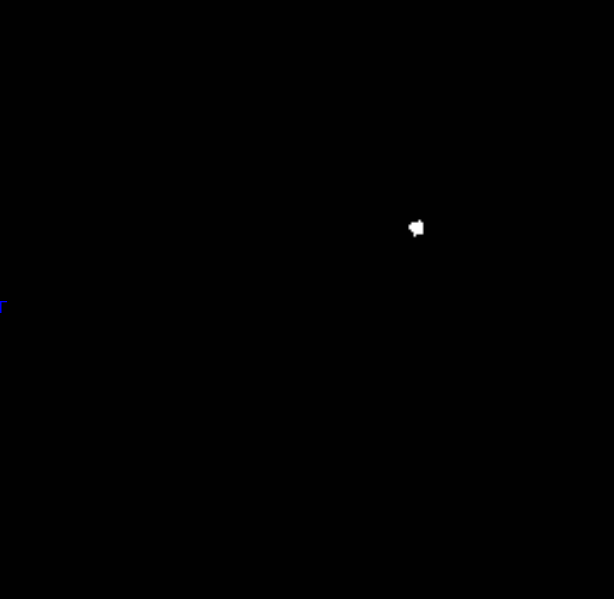}~~~
\includegraphics[width=0.13\textwidth]{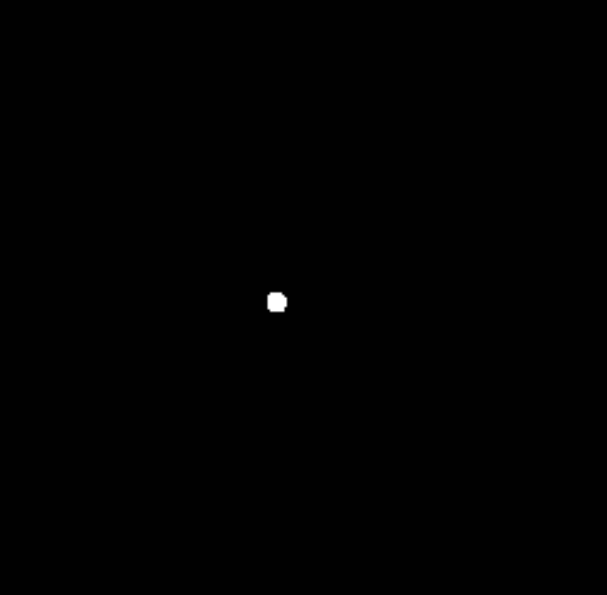}~~~
\includegraphics[width=0.129\textwidth]{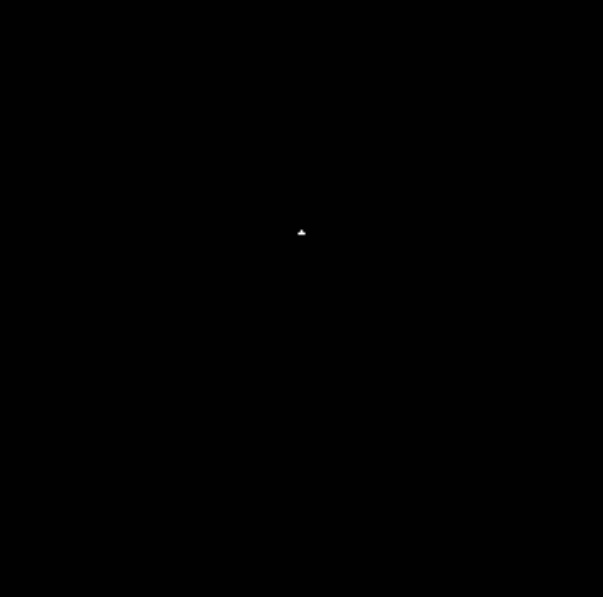}~~~
\includegraphics[width=0.13\textwidth]{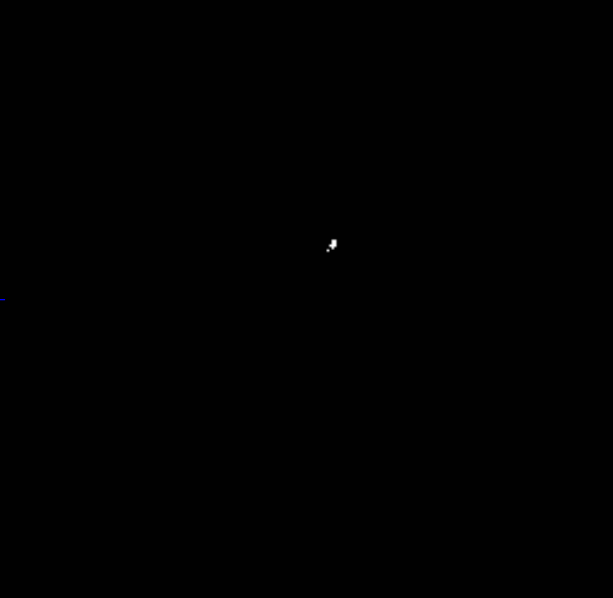}~~~
\includegraphics[width=0.13\textwidth]{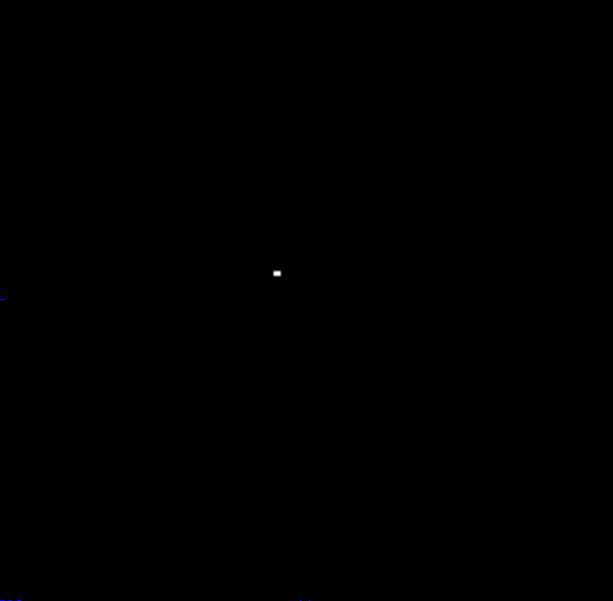}~~~
\includegraphics[width=0.13\textwidth]{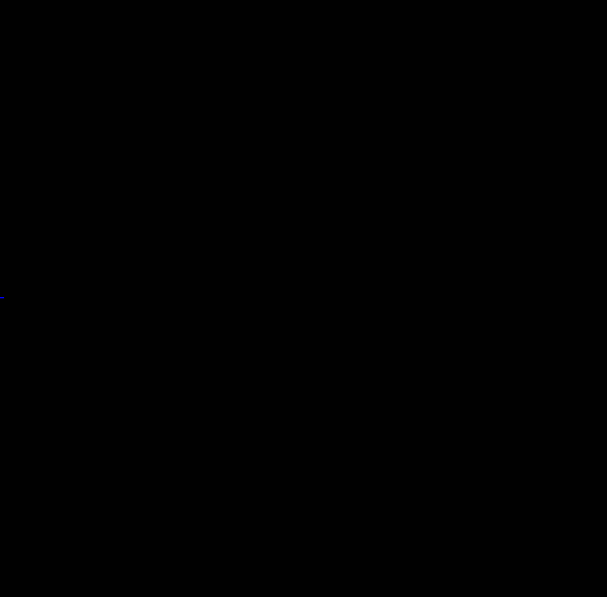}\\
\hspace{2pt}Pred:~
\includegraphics[width=0.13\textwidth]{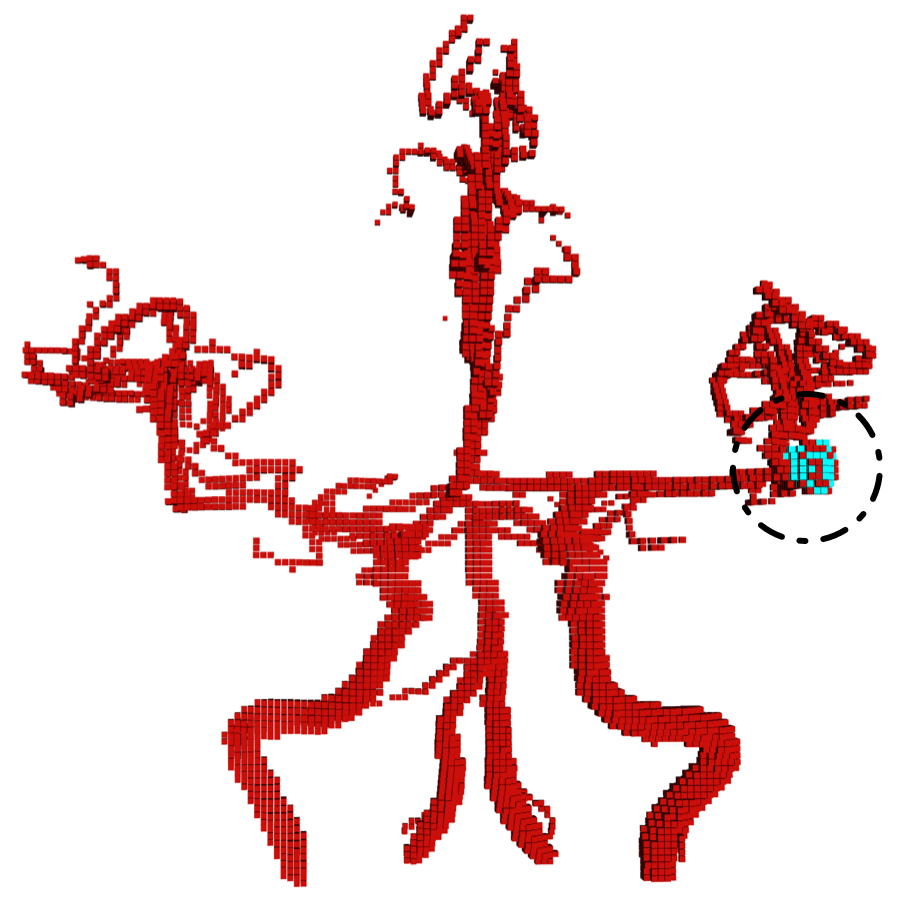}~~~
\includegraphics[width=0.13\textwidth]{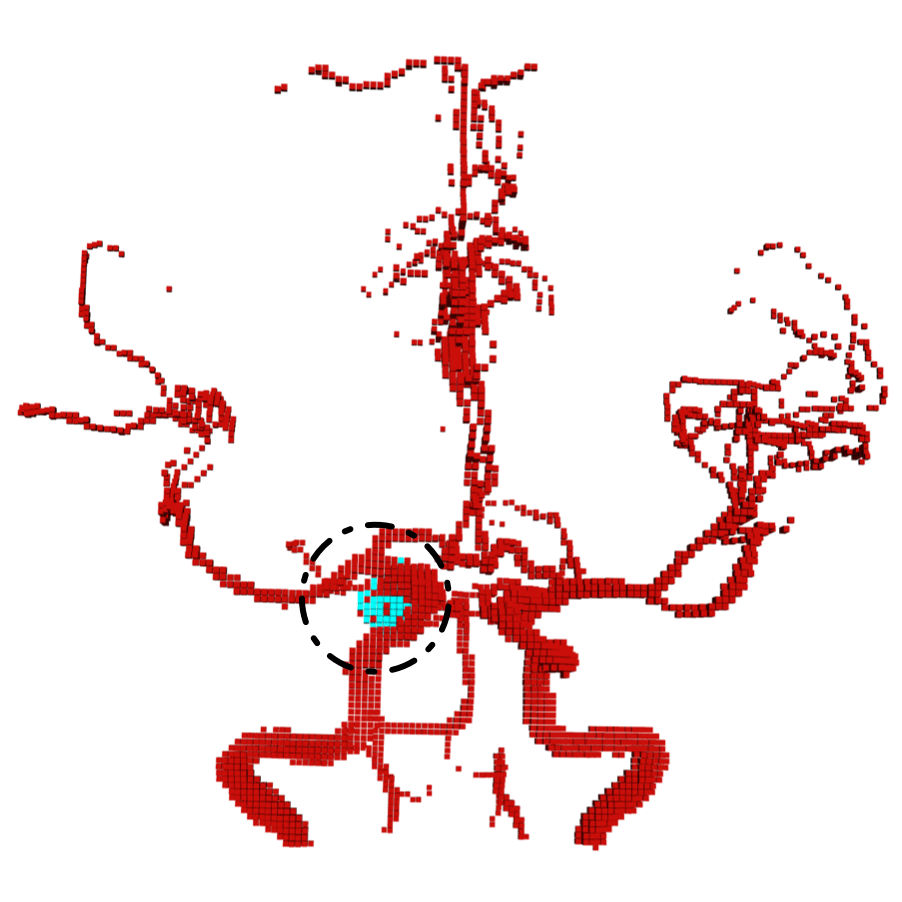}~~~
\includegraphics[width=0.13\textwidth]{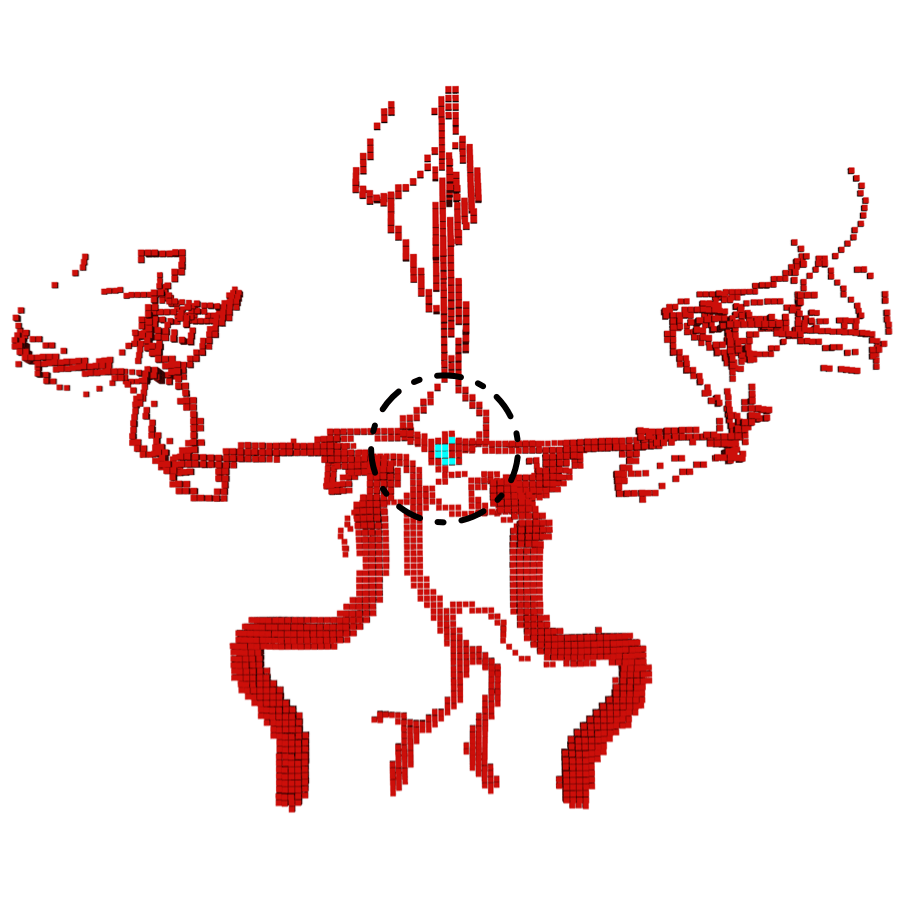}~~~
\includegraphics[width=0.13\textwidth]{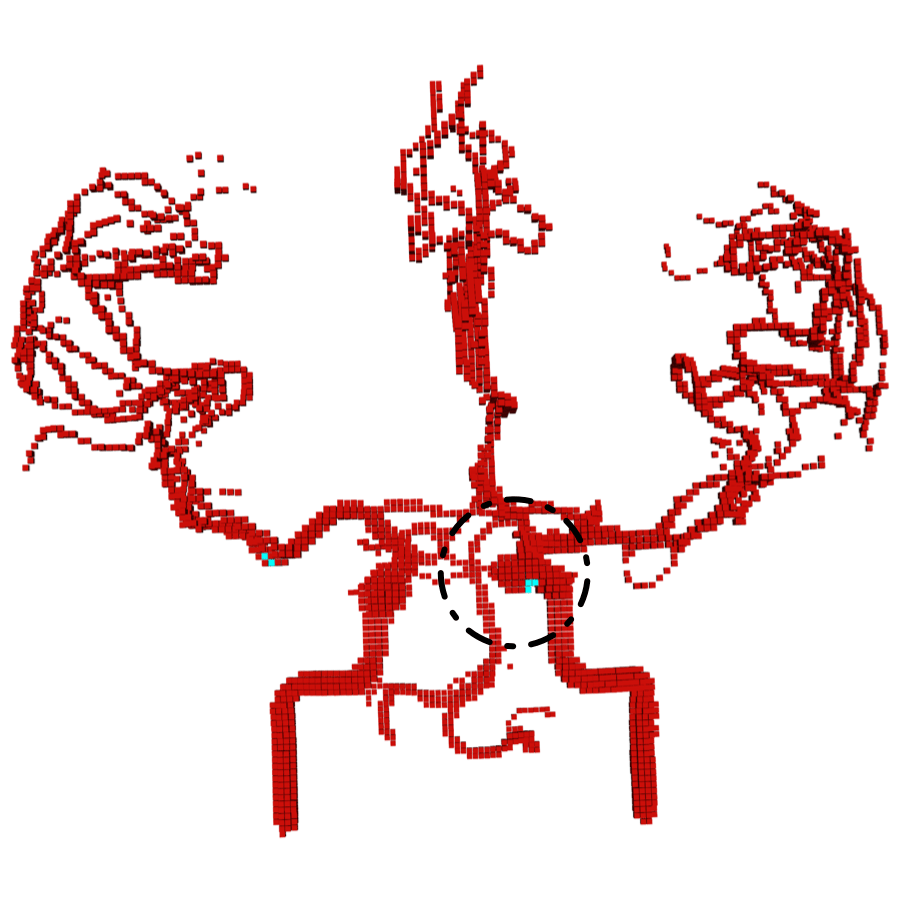}~~~
\includegraphics[width=0.13\textwidth]{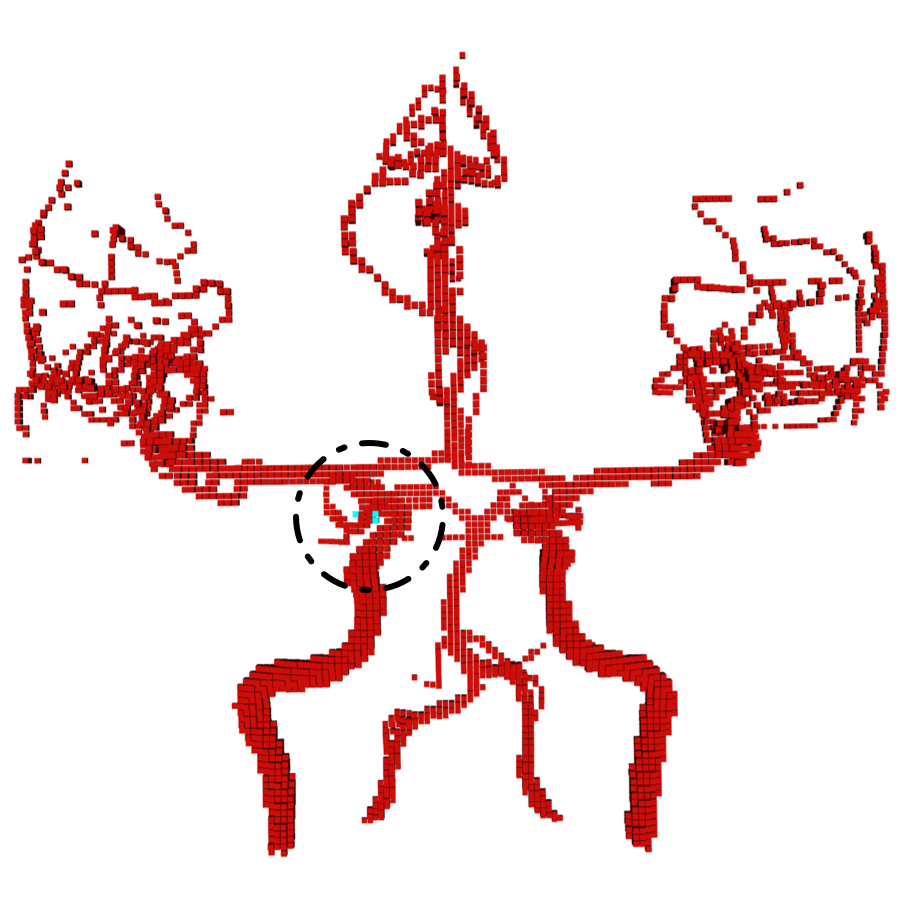}~~~
\includegraphics[width=0.13\textwidth]{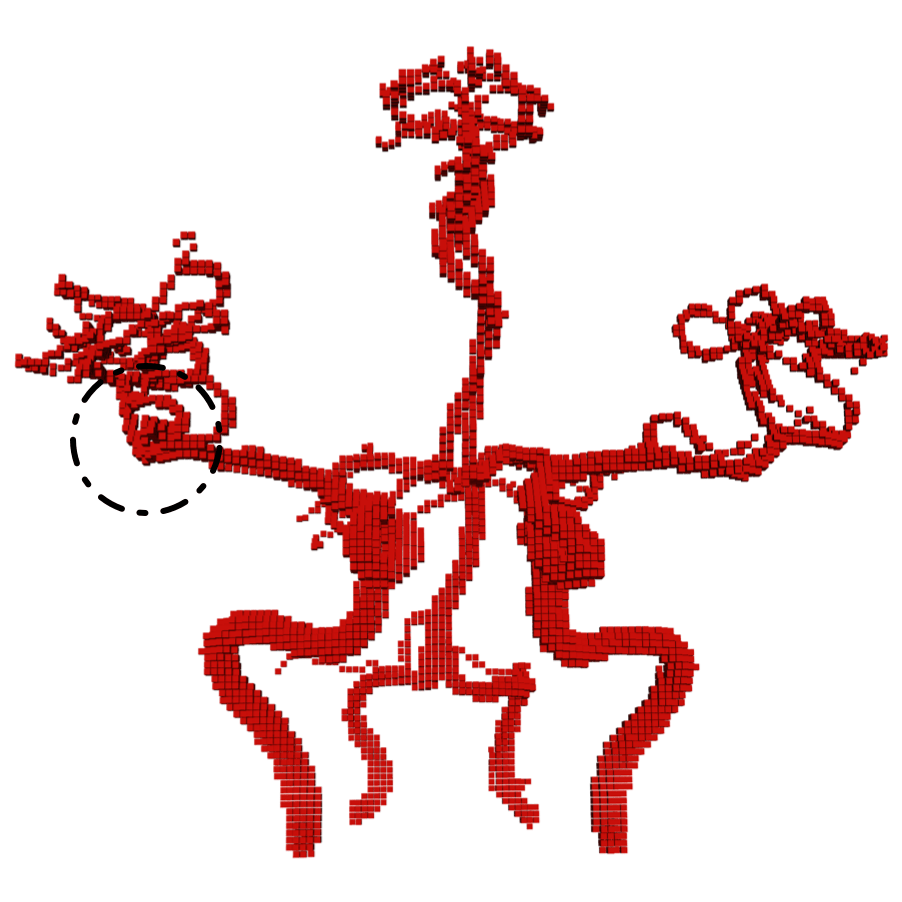}\\
\hspace{2pt}Pred:~
\begin{minipage}[t]{0.13\textwidth}
  \centering
  \raisebox{0.5\height}{\includegraphics[width=0.2\textwidth]{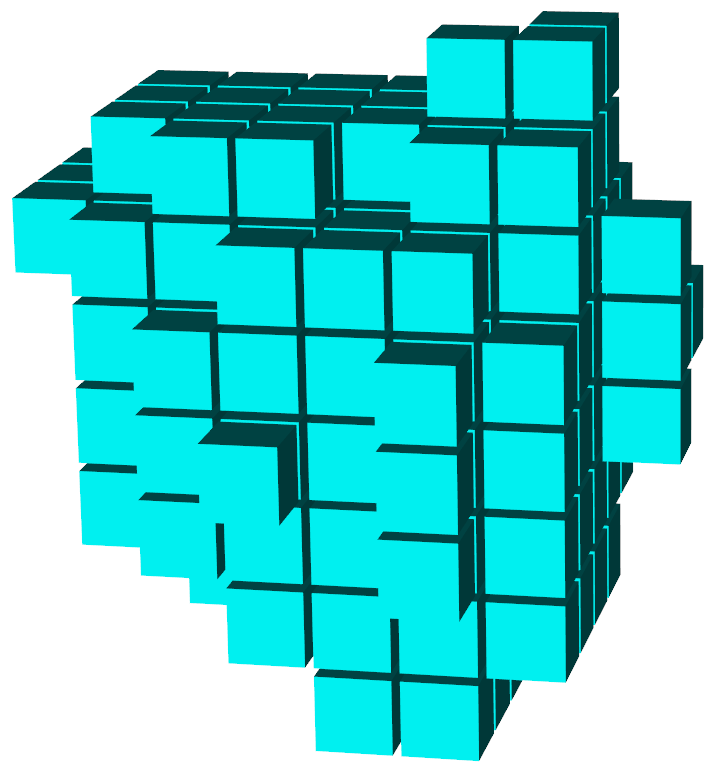}}
\end{minipage}~~~
\begin{minipage}[t]{0.13\textwidth}
  \centering
  \includegraphics[width=0.6\textwidth]{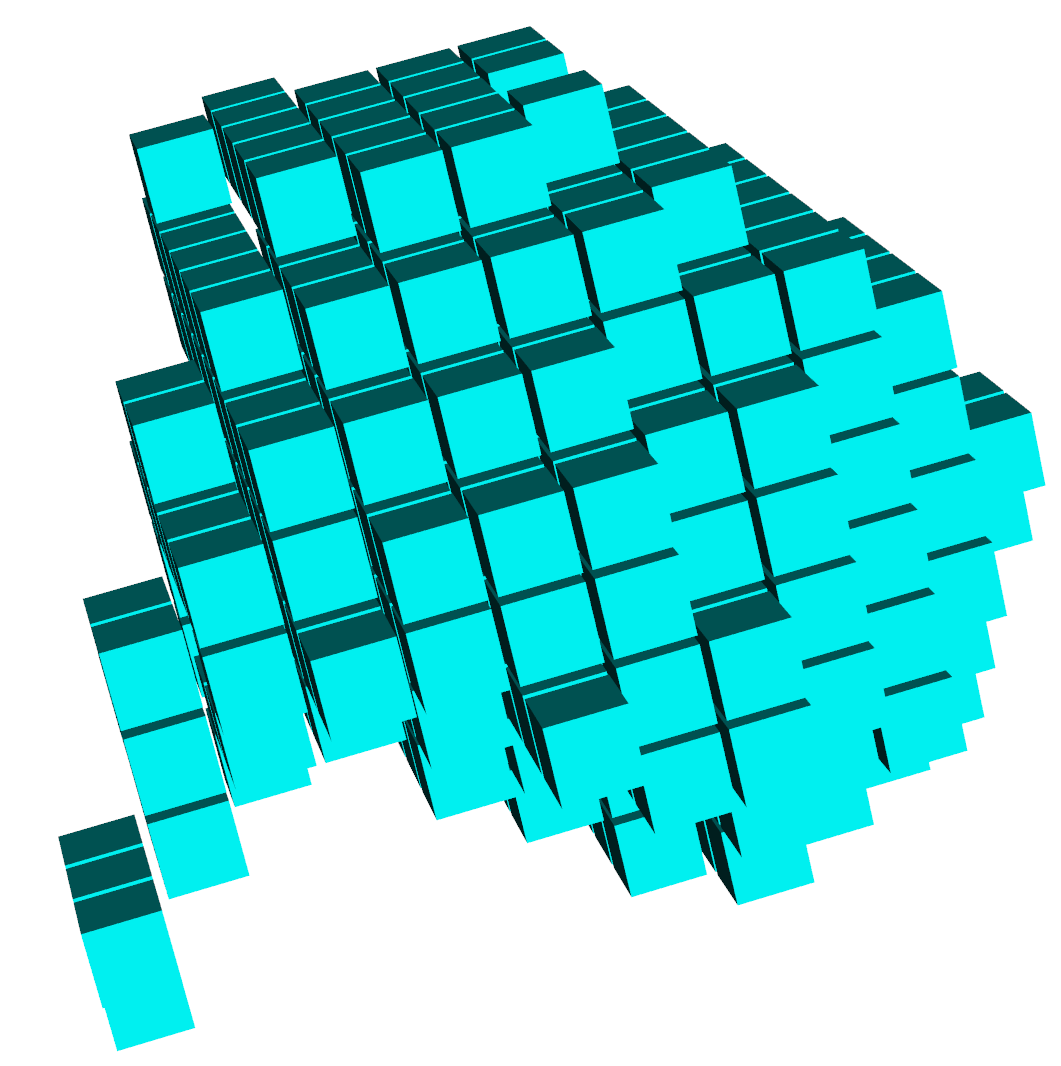}
\end{minipage}~~~
\begin{minipage}[t]{0.13\textwidth}
  \centering
  \raisebox{0.5\height}{\includegraphics[width=0.2\textwidth]{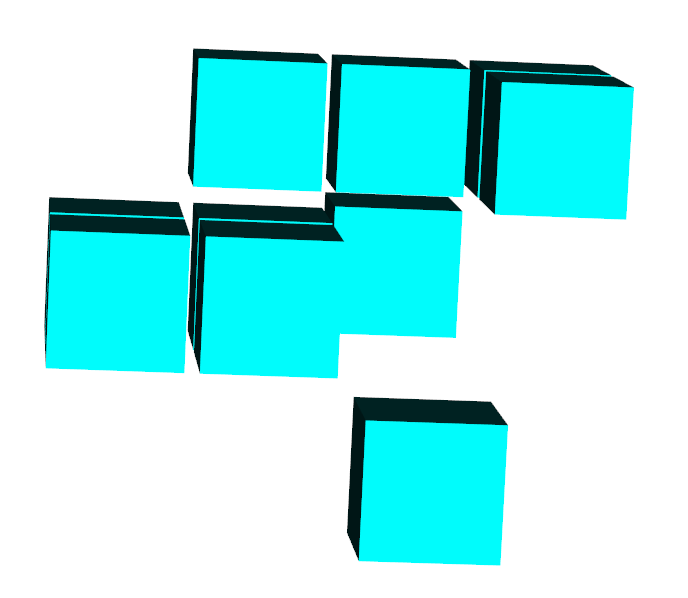}}
\end{minipage}~~~
\begin{minipage}[t]{0.13\textwidth}
  \centering
  \includegraphics[width=0.5\textwidth]{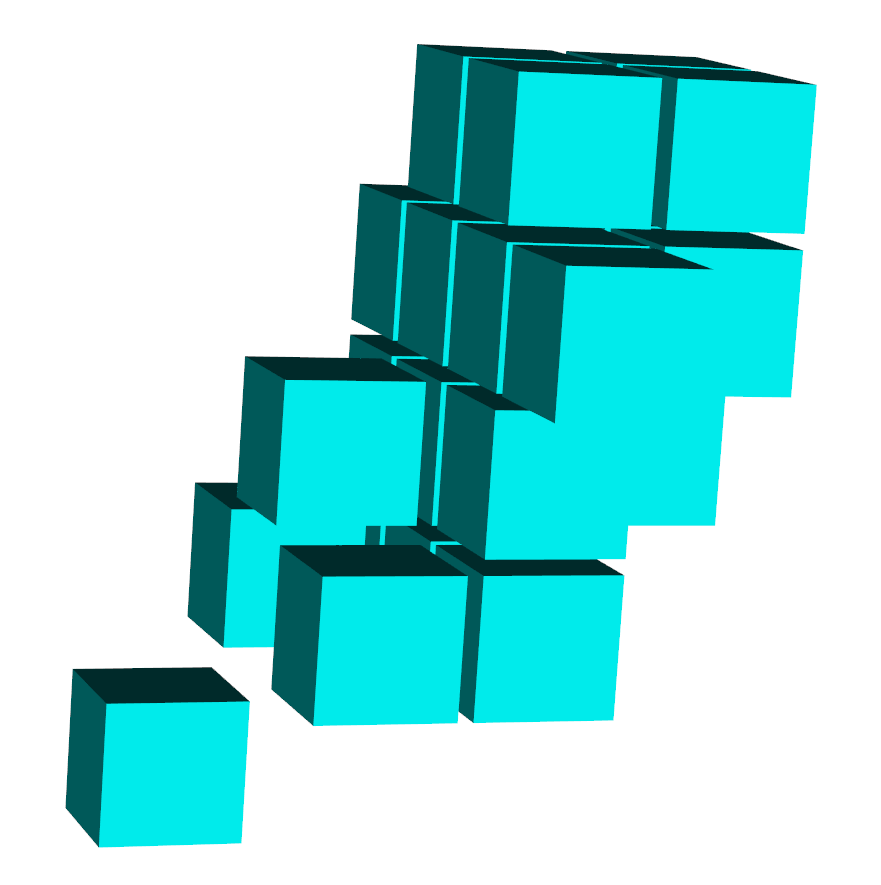}
\end{minipage}~~~
\begin{minipage}[t]{0.13\textwidth}
  \centering
  \raisebox{0.5\height}{\includegraphics[width=0.2\textwidth]{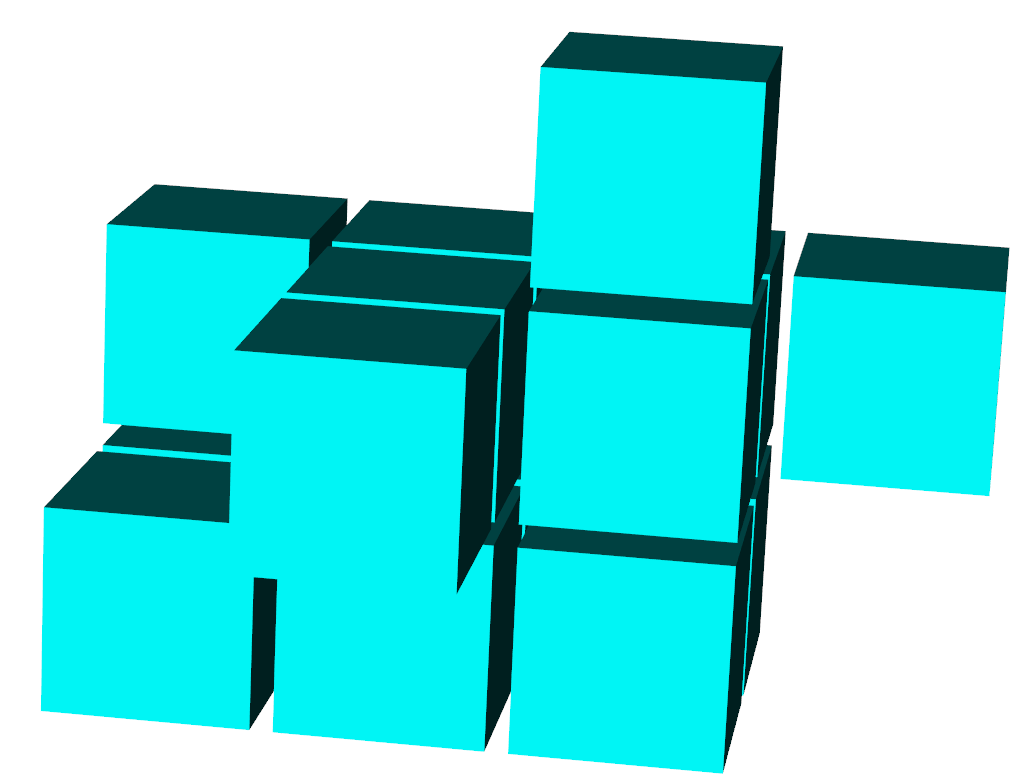}}
\end{minipage}~~~
\begin{minipage}[t]{0.13\textwidth}
  \centering
  \includegraphics[width=0.6\textwidth]{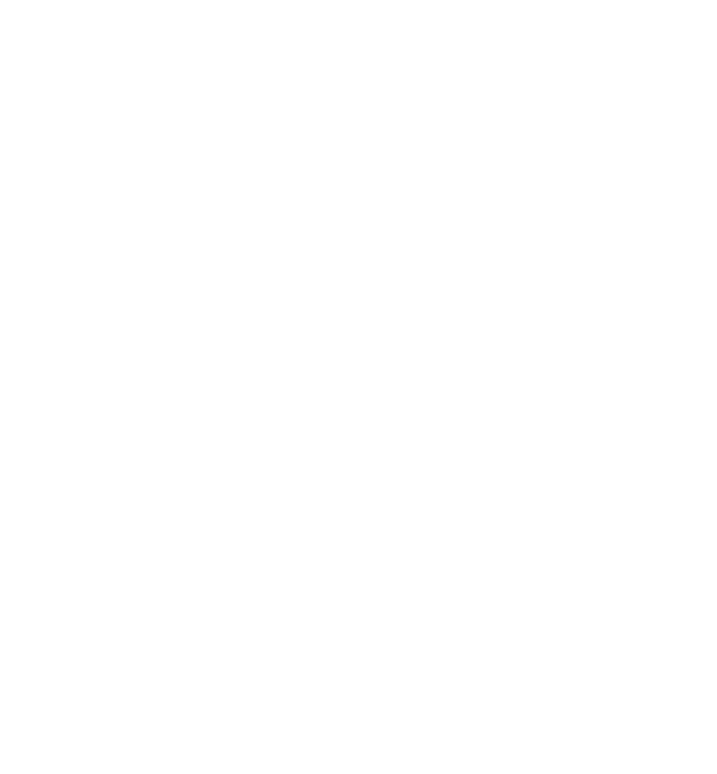}
\end{minipage}
%\\ {\phantom{x} \hspace{10pt} GT \hspace{45pt} Prediction \hspace{30pt} GT \& Prediction \hspace{55pt} Input \hspace{50pt} GT \hspace{50pt} Prediction}
%\vspace{-1mm}
\caption{Comparison of segmentation results. Both our method and DeepMedic yielded high segmentation accuracy on the two examples (the leftmost two columns); our method yielded significantly better results than DeepMedic (the middle three columns); both our method and DeepMedic could not obtain the parts with IA (the rightmost column), as the fragment with the IA was filtered out by our classification network. We also show the predicted results of DeepMedic by volume (the last two rows).} 
\label{fig:entrie-result}
\end{figure*}

%The segmentation results of the fragments are shown in Figure~\ref{fig:frag-result}. 
The DSC of the aneurysm parts was employed to evaluate the segmentation results. A comparison of the final segmentation results is shown in Figure~\ref{fig:entrie-result} and Table~\ref{tab:comparison}. 
The performance of the voxel-based method was comparable to that reported in the original paper~(\cite{sichtermann2019deep}).
Our surface-based method obtained much better segmentation results than the voxel-based method on most of the data. However, a few samples with tiny aneurysms were challenging both for the voxel-based method and for ours.

\subsubsection{DeepMedic with surface mask (DeepMedic S)}

\begin{figure}[th]
\centering
\includegraphics[width=0.32\linewidth, angle=180]{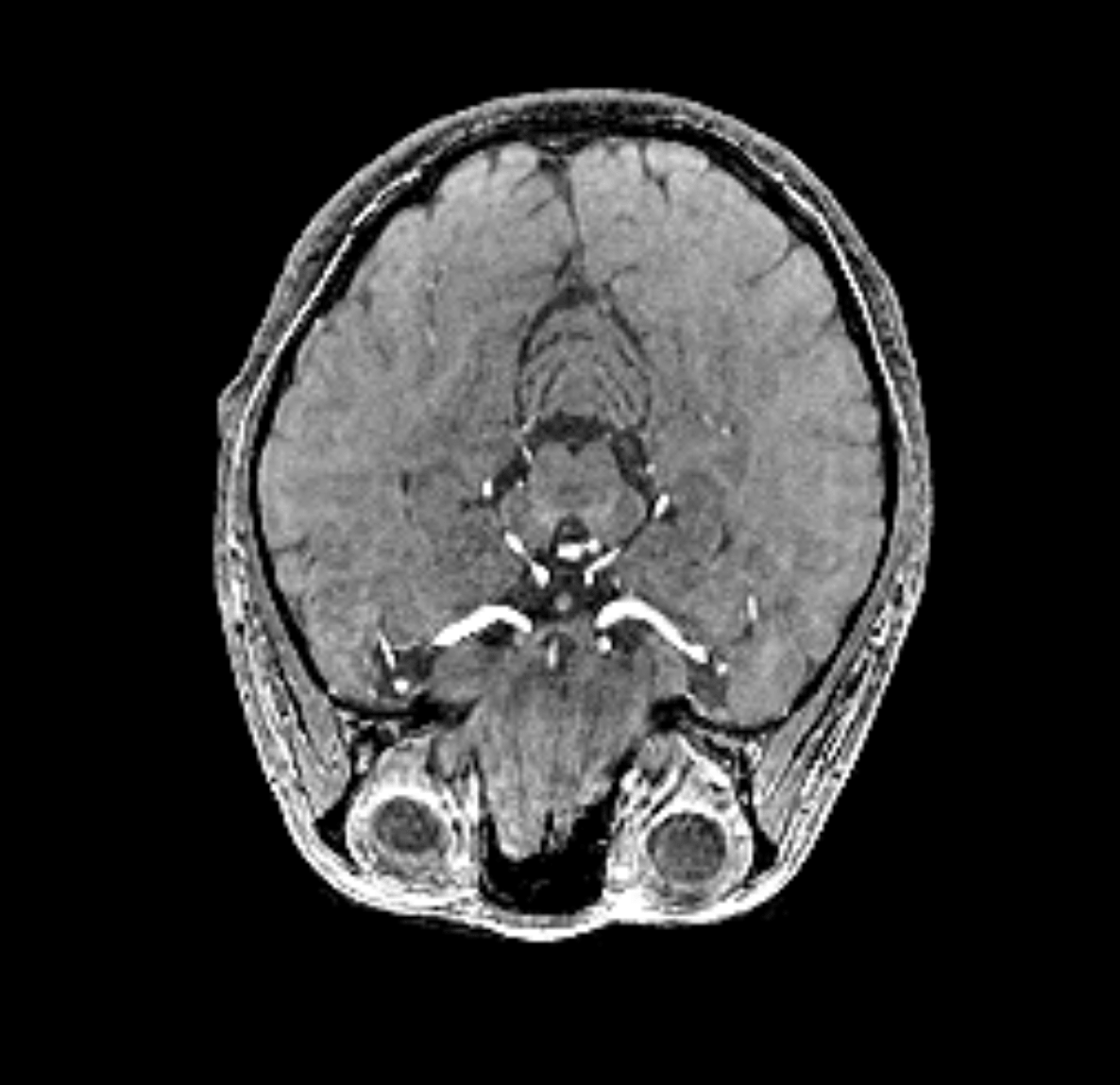}~
\includegraphics[width=0.32\linewidth, angle=180]{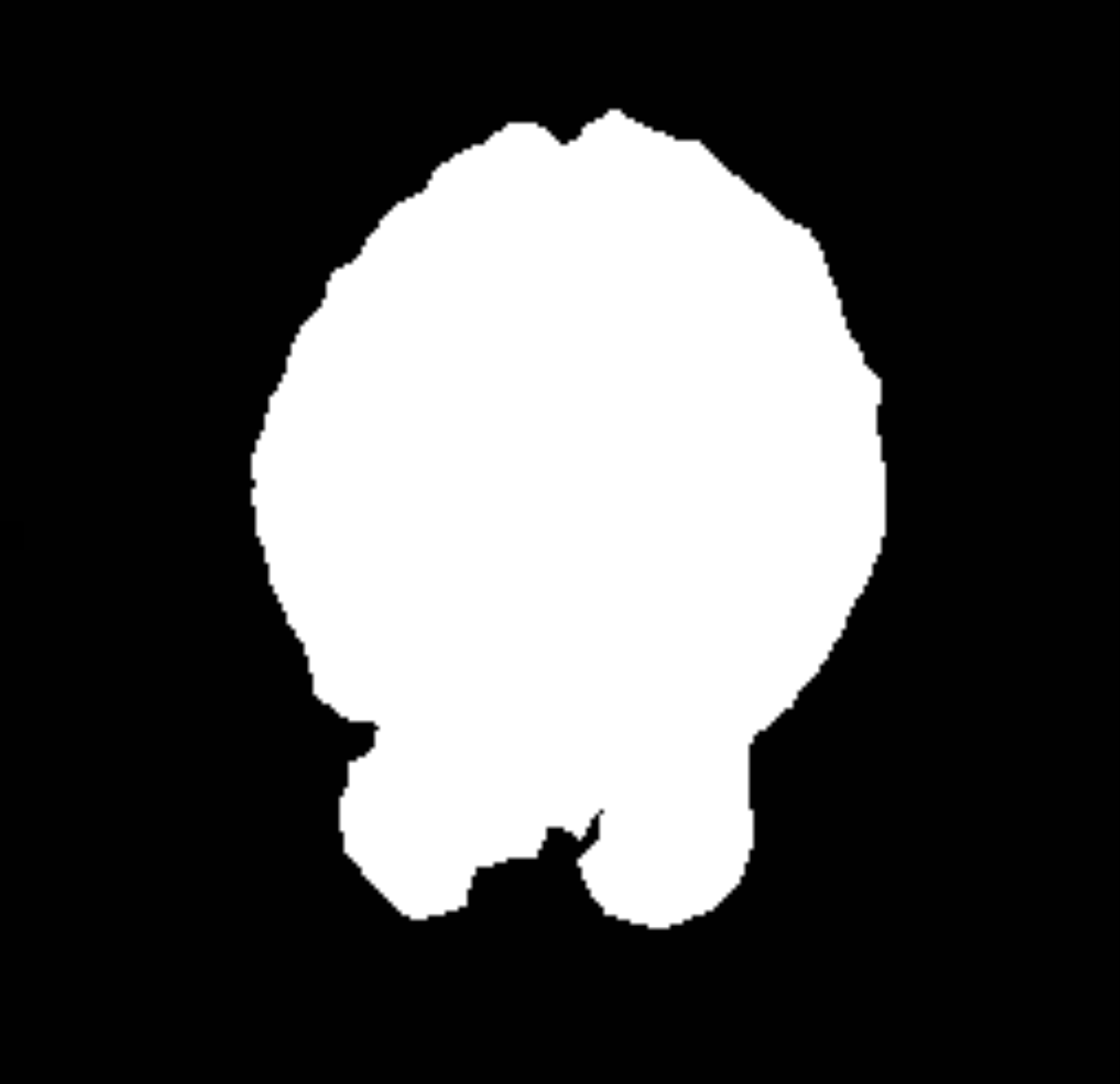}~
\includegraphics[width=0.32\linewidth, angle=180]{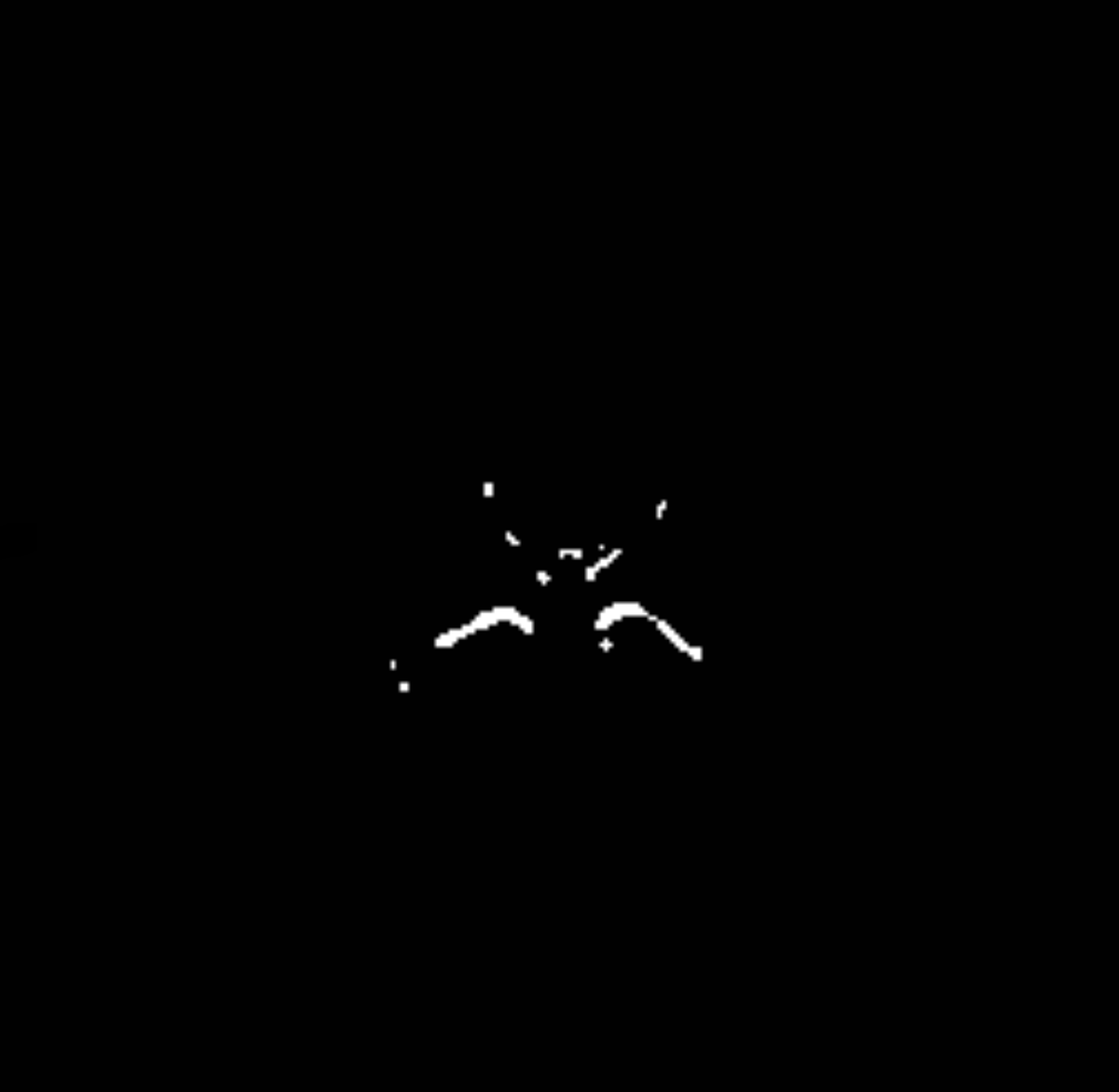}
\caption{Left: a slice of the original MRA images. Middle: the corresponding ROI mask of skull-stripping generated by BET2. Right: the corresponding ROI surface mask generated by our surface model to provide the same input region for the DeepMedic model.}
\label{fig:masks}
\end{figure}

To directly compare performance between voxel-based and points-based networks, we generated artery region masks by converting our entire surface models into solid models and then mapping them back to the original MRA images, as shown in Figure~\ref{fig:masks}. Thus, the voxel-based network obtained the same Region of Interest (ROI) as point-based models. We can see that the segmentation results were improved compared to the model trained with skull-stripping masks. However, the performance was still worse than that of our surface-based method. This comparative experiment shows that point-based networks can learn more accurate topological and geometric shape information compared to voxel-based models. 

\begin{table}[t]
\caption{Comparison of segmentation results in DSCs ($\%$).}
\label{tab:comparison}
% \centering
\setlength{\tabcolsep}{3pt}
\small
\begin{tabular}{c|c|c|c|c|c|c|c|c}
\hline
& \multicolumn{2}{c|}{Ours} & \multicolumn{2}{c|}{DeepMedic S} & \multicolumn{2}{c|}{DeepMedic B} & \multicolumn{2}{c}{3D U-Net} \\
& \multicolumn{2}{c|}{(Surface-based)} & \multicolumn{2}{c|}{(Voxel-based)} & \multicolumn{2}{c|}{(Voxel-based)} & \multicolumn{2}{c}{(Voxel-based)} \\
\hline
& Mean & STD & Mean & STD & Mean & STD & Mean & STD \\ \hline
Overall & \textbf{71.79} & 29.91 & 52.55 & 31.37 & 45.90 & 31.00 & - & - \\ \hline
Fold 0 & \textbf{73.10} & 32.55 & 59.38 & 28.20 & 56.56 & 28.77 & - & - \\ \hline
Fold 1 & \textbf{75.72} & 27.23 & 61.70 & 30.73 & 47.40 & 30.25 & - & - \\ \hline
Fold 2 & \textbf{73.81} & 30.43 & 47.96 & 32.34 & 38.86 & 31.57 & - & - \\ \hline
Fold 3 & \textbf{72.37} & 30.50 & 53.97 & 27.58 & 45.51 & 30.88 & - & - \\ \hline
Fold 4 & \textbf{64.17} & 34.85 & 40.26 & 35.27 & 41.20 & 33.22 & - & - \\ \hline
\end{tabular}
\end{table}

\section{Limitation}
Our current pipeline requires manual effort by medical experts to obtain surface models of intracranial artery networks. Thus, a possible criticism of our method is that this process severely limits its practical value. There are three reasons why we still believe that our method has significant practical value. First, neurosurgeons are presently already constructing surface models regularly in practice for preoperative examinations. Thus, we can assume that the surface model is already given in context.
%One might criticize that the most difficult part is already solved in this manual process and automatic segmentation presented in this paper is trivial. There are two possible counter arguments. 
Second, the construction of the surface model is mostly performed through simple thresholding~\cite{kin2012new}. An expert manually sets a threshold, and voxels with intensities higher than the threshold are automatically extracted. In this process, the expert does not pay attention to the details of individual aneurysms. Aneurysms need to be carefully segmented manually using surface editing tools in current practice, and we expect automation of this process to be highly appreciated. Finally, we expect that, with the advances in deep learning methods, surface extraction will become largely or even fully automatic in the future. Consequently, the entire process may be fully automated, which has the potential to significantly impact the field.

Our experimental results show the baseline performance of the proposed framework. We believe that our results can be further improved significantly by adjusting hyper-parameters.

\section{Conclusion}
In this study, we proposed a new surface-based framework for the segmentation of intracranial aneurysms from TOF-MRA images. Our framework applied a two-step design, classification-segmentation, using state-of-the-art point-based deep learning networks. We also designed sampling and refinement methods for the IA segmentation task. The segmentation results show that our framework significantly outperforms the existing voxel-based method.
% that the significant performance improvement is obtained by using our framework compared to an existing volume-based method.
Surface-based methods are as yet not prevalent in medical diagnosis and surgical planning. Our results show that surface-based methods can be a reliable alternative to popular voxel-based methods, and we hope this work may inspire further research efforts in this direction in other medical application domains.

%%
%% The acknowledgments section is defined using the "acks" environment
%% (and NOT an unnumbered section). This ensures the proper
%% identification of the section in the article metadata, and the
%% consistent spelling of the heading.
\begin{acks}
This research was supported by AMED under Grant Number JP18he1602001.
\end{acks}

\clearpage

%%
%% The next two lines define the bibliography style to be used, and
%% the bibliography file.
\bibliographystyle{ACM-Reference-Format}
\bibliography{sample-base}

%%
%% If your work has an appendix, this is the place to put it.
% \appendix

% \section{Research Methods}

% \subsection{Part One}

% Lorem ipsum dolor sit amet, consectetur adipiscing elit. Morbi
% malesuada, quam in pulvinar varius, metus nunc fermentum urna, id
% sollicitudin purus odio sit amet enim. Aliquam ullamcorper eu ipsum
% vel mollis. Curabitur quis dictum nisl. Phasellus vel semper risus, et
% lacinia dolor. Integer ultricies commodo sem nec semper.

% \subsection{Part Two}

% Etiam commodo feugiat nisl pulvinar pellentesque. Etiam auctor sodales
% ligula, non varius nibh pulvinar semper. Suspendisse nec lectus non
% ipsum convallis congue hendrerit vitae sapien. Donec at laoreet
% eros. Vivamus non purus placerat, scelerisque diam eu, cursus
% ante. Etiam aliquam tortor auctor efficitur mattis.

% \section{Online Resources}

% Nam id fermentum dui. Suspendisse sagittis tortor a nulla mollis, in
% pulvinar ex pretium. Sed interdum orci quis metus euismod, et sagittis
% enim maximus. Vestibulum gravida massa ut felis suscipit
% congue. Quisque mattis elit a risus ultrices commodo venenatis eget
% dui. Etiam sagittis eleifend elementum.

% Nam interdum magna at lectus dignissim, ac dignissim lorem
% rhoncus. Maecenas eu arcu ac neque placerat aliquam. Nunc pulvinar
% massa et mattis lacinia.

\end{document}